\documentclass[%
    reprint,
    superscriptaddress,
    amsmath,amssymb,
    aps,
    prl,
    noeprint,
    floatfix
]{revtex4-2}

\usepackage{graphicx}
\usepackage{dcolumn}
\usepackage{bm}
\usepackage{color}
\usepackage{multirow}
\usepackage{hyperref}
\usepackage[caption=false]{subfig}

\usepackage{xspace}

\graphicspath{{./}{./figures/}}

\newcommand{\eg}{\emph{e.g.}\xspace}

\begin{document}

\title{Observation of Seven Astrophysical Tau Neutrino Candidates with IceCube}

\affiliation{III. Physikalisches Institut, RWTH Aachen University, D-52056 Aachen, Germany}
\affiliation{Department of Physics, University of Adelaide, Adelaide, 5005, Australia}
\affiliation{Dept. of Physics and Astronomy, University of Alaska Anchorage, 3211 Providence Dr., Anchorage, AK 99508, USA}
\affiliation{Dept. of Physics, University of Texas at Arlington, 502 Yates St., Science Hall Rm 108, Box 19059, Arlington, TX 76019, USA}
\affiliation{CTSPS, Clark-Atlanta University, Atlanta, GA 30314, USA}
\affiliation{School of Physics and Center for Relativistic Astrophysics, Georgia Institute of Technology, Atlanta, GA 30332, USA}
\affiliation{Dept. of Physics, Southern University, Baton Rouge, LA 70813, USA}
\affiliation{Dept. of Physics, University of California, Berkeley, CA 94720, USA}
\affiliation{Lawrence Berkeley National Laboratory, Berkeley, CA 94720, USA}
\affiliation{Institut f{\"u}r Physik, Humboldt-Universit{\"a}t zu Berlin, D-12489 Berlin, Germany}
\affiliation{Fakult{\"a}t f{\"u}r Physik {\&} Astronomie, Ruhr-Universit{\"a}t Bochum, D-44780 Bochum, Germany}
\affiliation{Universit{\'e} Libre de Bruxelles, Science Faculty CP230, B-1050 Brussels, Belgium}
\affiliation{Vrije Universiteit Brussel (VUB), Dienst ELEM, B-1050 Brussels, Belgium}
\affiliation{Department of Physics and Laboratory for Particle Physics and Cosmology, Harvard University, Cambridge, MA 02138, USA}
\affiliation{Dept. of Physics, Massachusetts Institute of Technology, Cambridge, MA 02139, USA}
\affiliation{Dept. of Physics and The International Center for Hadron Astrophysics, Chiba University, Chiba 263-8522, Japan}
\affiliation{Department of Physics, Loyola University Chicago, Chicago, IL 60660, USA}
\affiliation{Dept. of Physics and Astronomy, University of Canterbury, Private Bag 4800, Christchurch, New Zealand}
\affiliation{Dept. of Physics, University of Maryland, College Park, MD 20742, USA}
\affiliation{Dept. of Astronomy, Ohio State University, Columbus, OH 43210, USA}
\affiliation{Dept. of Physics and Center for Cosmology and Astro-Particle Physics, Ohio State University, Columbus, OH 43210, USA}
\affiliation{Niels Bohr Institute, University of Copenhagen, DK-2100 Copenhagen, Denmark}
\affiliation{Dept. of Physics, TU Dortmund University, D-44221 Dortmund, Germany}
\affiliation{Dept. of Physics and Astronomy, Michigan State University, East Lansing, MI 48824, USA}
\affiliation{Dept. of Physics, University of Alberta, Edmonton, Alberta, Canada T6G 2E1}
\affiliation{Erlangen Centre for Astroparticle Physics, Friedrich-Alexander-Universit{\"a}t Erlangen-N{\"u}rnberg, D-91058 Erlangen, Germany}
\affiliation{Physik-department, Technische Universit{\"a}t M{\"u}nchen, D-85748 Garching, Germany}
\affiliation{D{\'e}partement de physique nucl{\'e}aire et corpusculaire, Universit{\'e} de Gen{\`e}ve, CH-1211 Gen{\`e}ve, Switzerland}
\affiliation{Dept. of Physics and Astronomy, University of Gent, B-9000 Gent, Belgium}
\affiliation{Dept. of Physics and Astronomy, University of California, Irvine, CA 92697, USA}
\affiliation{Karlsruhe Institute of Technology, Institute for Astroparticle Physics, D-76021 Karlsruhe, Germany }
\affiliation{Karlsruhe Institute of Technology, Institute of Experimental Particle Physics, D-76021 Karlsruhe, Germany }
\affiliation{Dept. of Physics, Engineering Physics, and Astronomy, Queen's University, Kingston, ON K7L 3N6, Canada}
\affiliation{Department of Physics {\&} Astronomy, University of Nevada, Las Vegas, NV, 89154, USA}
\affiliation{Nevada Center for Astrophysics, University of Nevada, Las Vegas, NV 89154, USA}
\affiliation{Dept. of Physics and Astronomy, University of Kansas, Lawrence, KS 66045, USA}
\affiliation{Centre for Cosmology, Particle Physics and Phenomenology - CP3, Universit{\'e} catholique de Louvain, Louvain-la-Neuve, Belgium}
\affiliation{Department of Physics, Mercer University, Macon, GA 31207-0001, USA}
\affiliation{Dept. of Astronomy, University of Wisconsin{\textendash}Madison, Madison, WI 53706, USA}
\affiliation{Dept. of Physics and Wisconsin IceCube Particle Astrophysics Center, University of Wisconsin{\textendash}Madison, Madison, WI 53706, USA}
\affiliation{Institute of Physics, University of Mainz, Staudinger Weg 7, D-55099 Mainz, Germany}
\affiliation{Department of Physics, Marquette University, Milwaukee, WI, 53201, USA}
\affiliation{Institut f{\"u}r Kernphysik, Westf{\"a}lische Wilhelms-Universit{\"a}t M{\"u}nster, D-48149 M{\"u}nster, Germany}
\affiliation{Bartol Research Institute and Dept. of Physics and Astronomy, University of Delaware, Newark, DE 19716, USA}
\affiliation{Dept. of Physics, Yale University, New Haven, CT 06520, USA}
\affiliation{Columbia Astrophysics and Nevis Laboratories, Columbia University, New York, NY 10027, USA}
\affiliation{Dept. of Physics, University of Oxford, Parks Road, Oxford OX1 3PU, United Kingdom}
\affiliation{Dipartimento di Fisica e Astronomia Galileo Galilei, Universit{\`a} Degli Studi di Padova, 35122 Padova PD, Italy}
\affiliation{Dept. of Physics, Drexel University, 3141 Chestnut Street, Philadelphia, PA 19104, USA}
\affiliation{Physics Department, South Dakota School of Mines and Technology, Rapid City, SD 57701, USA}
\affiliation{Dept. of Physics, University of Wisconsin, River Falls, WI 54022, USA}
\affiliation{Dept. of Physics and Astronomy, University of Rochester, Rochester, NY 14627, USA}
\affiliation{Department of Physics and Astronomy, University of Utah, Salt Lake City, UT 84112, USA}
\affiliation{Oskar Klein Centre and Dept. of Physics, Stockholm University, SE-10691 Stockholm, Sweden}
\affiliation{Dept. of Physics and Astronomy, Stony Brook University, Stony Brook, NY 11794-3800, USA}
\affiliation{Dept. of Physics, Sungkyunkwan University, Suwon 16419, Korea}
\affiliation{Institute of Physics, Academia Sinica, Taipei, 11529, Taiwan}
\affiliation{Dept. of Physics and Astronomy, University of Alabama, Tuscaloosa, AL 35487, USA}
\affiliation{Dept. of Astronomy and Astrophysics, Pennsylvania State University, University Park, PA 16802, USA}
\affiliation{Dept. of Physics, Pennsylvania State University, University Park, PA 16802, USA}
\affiliation{Dept. of Physics and Astronomy, Uppsala University, Box 516, S-75120 Uppsala, Sweden}
\affiliation{Dept. of Physics, University of Wuppertal, D-42119 Wuppertal, Germany}
\affiliation{Deutsches Elektronen-Synchrotron DESY, Platanenallee 6, 15738 Zeuthen, Germany }

\author{R. Abbasi}
\affiliation{Department of Physics, Loyola University Chicago, Chicago, IL 60660, USA}
\author{M. Ackermann}
\affiliation{Deutsches Elektronen-Synchrotron DESY, Platanenallee 6, 15738 Zeuthen, Germany }
\author{J. Adams}
\affiliation{Dept. of Physics and Astronomy, University of Canterbury, Private Bag 4800, Christchurch, New Zealand}
\author{S. K. Agarwalla}
\thanks{also at Institute of Physics, Sachivalaya Marg, Sainik School Post, Bhubaneswar 751005, India}
\affiliation{Dept. of Physics and Wisconsin IceCube Particle Astrophysics Center, University of Wisconsin{\textendash}Madison, Madison, WI 53706, USA}
\author{J. A. Aguilar}
\affiliation{Universit{\'e} Libre de Bruxelles, Science Faculty CP230, B-1050 Brussels, Belgium}
\author{M. Ahlers}
\affiliation{Niels Bohr Institute, University of Copenhagen, DK-2100 Copenhagen, Denmark}
\author{J.M. Alameddine}
\affiliation{Dept. of Physics, TU Dortmund University, D-44221 Dortmund, Germany}
\author{N. M. Amin}
\affiliation{Bartol Research Institute and Dept. of Physics and Astronomy, University of Delaware, Newark, DE 19716, USA}
\author{K. Andeen}
\affiliation{Department of Physics, Marquette University, Milwaukee, WI, 53201, USA}
\author{G. Anton}
\affiliation{Erlangen Centre for Astroparticle Physics, Friedrich-Alexander-Universit{\"a}t Erlangen-N{\"u}rnberg, D-91058 Erlangen, Germany}
% ...

\collaboration{IceCube Collaboration}
\email{analysis@icecube.wisc.edu}

\begin{abstract}

\noindent We report on a measurement of astrophysical tau neutrinos with 9.7 years of IceCube data. Using convolutional neural networks trained on images derived from simulated events, seven candidate $\nu_\tau$ events were found with visible energies ranging from roughly 20~TeV to 1~PeV and a median expected parent $\nu_\tau$ energy of about 200~TeV. 
Considering backgrounds from astrophysical and atmospheric neutrinos, and muons from $\pi^\pm/K^\pm$ decays in atmospheric air showers, we obtain a total estimated background of about 0.5 events, dominated by non-$\nu_\tau$ astrophysical neutrinos.
Thus, we rule out the absence of astrophysical $\nu_\tau$ at the $5\sigma$ level. The measured astrophysical $\nu_\tau$ flux is consistent with expectations based on previously published IceCube astrophysical neutrino flux measurements and neutrino oscillations.

\end{abstract}

\setcounter{page}{1}

\twocolumngrid

\maketitle

\noindent
In 2013 IceCube discovered a flux of neutrinos of astrophysical origin~\cite{IceCube:2013cdw,HESE2y,HESE3y}. 
The astrophysical neutrino ($\nu^\mathrm{astro}$) flux normalization and index $\gamma$ carry information about neutrino sources and their environments~\cite{Fermi:1949ee,Bell-1978,Gaisser1990,Protheroe,KazEll,Begelman,Stecketal,StecketalE,MannBier,Matthews:2020lig,Winter2013,murase}. 
Different $\nu^\mathrm{astro}$ production mechanisms lead to different $\nu_e\!:\!\nu_\mu\!:\!\nu_\tau$ ratios at the sources but, after standard neutrino oscillations over astrophysical distances, detectable numbers of all three neutrino flavors are expected at Earth~\cite{LearnedPakvasa,Athar2006,KashtiWaxman,Klein2013,Lipari2007,Bustamante2015,Esmaili2009,nufit,nufit2}.  Previous measurements at lower energies, using neutrinos produced at accelerators and in the atmosphere ($\nu^\mathrm{atm}$), have detected $\nu_\tau$ produced directly~\cite{DONuT:2007bsg} and through neutrino oscillations~\cite{OPERA:2019kzo,Super-Kamiokande:2017edb,IceCube:2019dqi}.  At the much higher energies accessible to this analysis, $\nu_\tau^\mathrm{atm}$ are strongly suppressed relative to $\nu_\tau^\mathrm{astro}$~\cite{Bhattacharya:2016jce}, while an unexpected level of presence of $\nu_\tau^\mathrm{astro}$ in the $\nu^\mathrm{astro}$ flux could be an indication of new physics~\cite{Arguelles:2015dca,Pagliaroli:2015rca,Shoemaker:2015qul,Brdar:2016thq,Bustamante:2016ciw,Klop:2017dim,Rasmussen:2017ert,Bustamante:2018mzu,Denton:2018aml,Farzan:2018pnk,IceCube:2021tdn,Abdullahi:2020rge,Abraham:2022jse,Ackermann:2022rqc}.  

Previous analyses~\cite{PhysRevD.86.022005,PhysRevD.93.022001,IceCube:2020fpi,Meier:2019ypu} by IceCube to detect $\nu_{\tau}^\mathrm{astro}$ included searches for double-cascade signatures, such as the distinctive ``double bang''~\cite{LearnedPakvasa} in the full detector or ``double pulse'' (DP) waveforms in one or two individual photosensors.  The DP signature is produced by the distinct arrival times of light signals at one or more photosensors from the $\nu_\tau$ interaction and $\tau$ decay vertices.  IceCube previously observed two candidate $\nu_{\tau}^\mathrm{astro}$,  ruling out the null hypothesis of no $\nu_{\tau}^\mathrm{astro}$ at $2.8\sigma$~\cite{IceCube:2020fpi}.  The analysis presented in this Letter reports on the low-background, high-significance detection of seven $\nu_{\tau}^\mathrm{astro}$ candidate events through the use of convolutional neural networks (CNNs).

IceCube~\cite{IceCube:2016zyt} is a neutrino observatory 
with 5160 Digital Optical Modules (DOMs) on 86 strings~\cite{IceCube:2016zyt,IceCube:2011ucd} 
in a cubic kilometer of ice at the South Pole. Charged particles produced in neutrino interactions emit Cherenkov light~\cite{Cherenkov:1934ilx} while propagating through the ice; photomultiplier tubes in the DOMs convert this light into electrical pulses that are digitized \emph{in situ}. Light is deposited in the detector in several distinct patterns: long tracks, single cascades, and double cascades.  Tracks are produced by muons from, \eg, $\nu_\mu$ charged-current (CC) interactions, and can start or end inside, or pass through, the detector. Single cascades arise from electromagnetic and/or hadronic particle showers produced by deep inelastic neutrino-nucleon interactions in or near the detector, or by $\overline{\nu}_e$ via the Glashow Resonance~\cite{gr,IceCube:2021rpz}.
Double cascades are formed by high-energy $\nu_{\tau,\,\mathrm{CC}}$ interactions in or near the detector that produce a hadronic shower and a $\tau$ lepton at the interaction vertex, followed by a second electromagnetic or hadronic shower at the $\tau$ decay vertex (BR[$\tau \to (e,h)] \simeq 83\%)$.
With a decay length of $\sim 50~\mathrm{m/PeV}$, the $\tau$ can travel a macroscopic distance in the ice.  For $\nu_\tau$ energies satisfying $E_{\nu_\tau} \gtrsim 1~\mathrm{PeV}$ and favorable geometric containment, the double-bang signature can be created, with two energetic and well-separated cascades.  Such events are intrinsically rare.  In contrast, for lower $E_{\nu_\tau}$ between roughly $50~\mathrm{TeV} - 1~\mathrm{PeV}$, the $\nu_\tau$ flux is expected to be higher but in CC interactions the two cascades are closer together.  Two cascades as close as about 10~m can produce distinctive patterns correlated across multiple DOMs and strings as the light from each cascade passes by, as well as DP waveforms in one or more DOMs.

\begin{figure}[h]
\centering
\includegraphics[width=0.95\columnwidth]{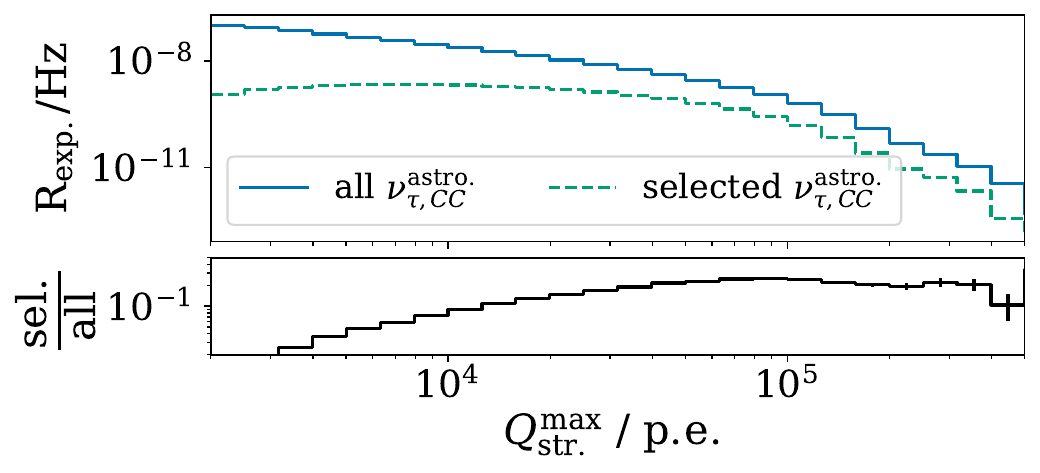}
\caption{Top: Simulated rate of $\nu_{\tau}^\mathrm{astro}$ CC events binned by the number of p.e.~detected by DOMs on the most illuminated string in the event, $Q_{\rm{str.}}^{\rm{max}}$, before any selection criteria (solid) and after the CNN-based criteria (dashed) described in the text. (Downward-going cosmic-ray muons trigger the detector at about 3~kHz, are effectively removed by our selection criteria, and are not shown on the plot; other backgrounds are similarly heavily reduced and also not shown.) Bottom: Ratio of rates (selected/all), showing that signal efficiency grows above about 2000~p.e. 
The IceCube ``GlobalFit'' $\nu^\mathrm{astro}$ flux~\protect\cite{IceCube:2015gsk} is assumed.  (Error bars statistical only.)}
\label{fig:qst0}
\end{figure}

We analyzed $9.7$ years of IceCube data from 2011--2020, triggering on approximately $10^{12}$ downward-going cosmic-ray muons, $10^{6}$ $\nu^\mathrm{atm}$, and $10^4$ $\nu^\mathrm{astro}$~\cite{IceCube:2015gsk,IceCube:2021uhz,IceCube:2018pgc,IceCube:2020wum}. We required that the DOMs on the most illuminated string collected at least 2000 photoelectrons (p.e.) (see Fig.~\ref{fig:qst0}) and at least 10~p.e.~in the two next-highest-charge, nearest-neighbor strings.  Signal events will appear more like cascades than tracks in IceCube, so we also selected events whose morphology was better described by the cascade hypothesis.  Aside from 0.6\% (22 live-days) of the data sample used to confirm agreement between data and simulation (data that were subsequently excluded from our analysis and which contained no signal-like events), we performed a ``blind'' analysis that only used simulated data to devise all selection criteria and analysis methodologies.  After application of these initial selection criteria, there was roughly 300 times more background than signal.  The expected number of p.e. on the most illuminated string for $\nu_{\tau}^\mathrm{astro}$ CC events after application of these criteria, and additional CNN criteria described below, is shown in Fig.~\ref{fig:qst0}.

\begin{figure}[h]
\centering
\includegraphics[width=0.98\columnwidth]{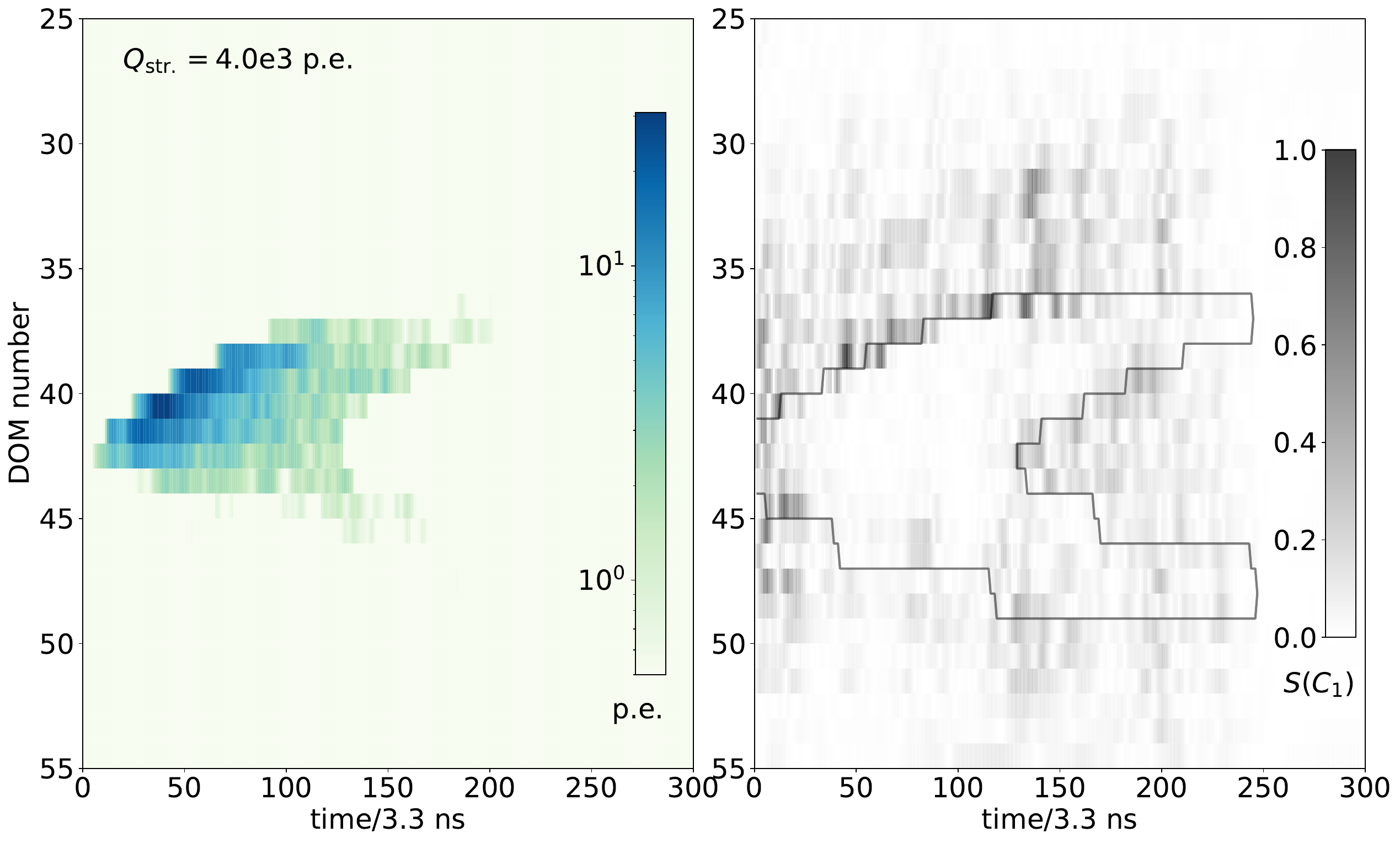}
\caption{Candidate $\nu_{\tau}^\mathrm{astro}$ detected in Sep. 2015.    
The left plot shows the DOM number (proportional to depth) versus the time of the digitized PMT signal in 3.3~ns bins for the highest-charge string, with the  scale giving the signal amplitude in p.e.~in each time bin. The total p.e.~detected on the string, $Q_\mathrm{str.}$, is shown.  The right plot shows $S(C_1)$, that string's saliency map for $C_1$, with 
darker regions indicating where the $C_1$ score is more sensitive to a changing light level (see text).  (The Appendix shows three-string views and signed saliencies for all seven $\nu_\tau^\mathrm{astro}$ candidates.)}
\label{fig:evt_display}
\end{figure}

We then created 2-d images of DOM number (corresponding to depth) vs. time in 3.3~ns bins, with each pixel's brightness proportional to the digitized waveform amplitude in that time bin.  Images were created for the 180 DOMs on the most illuminated string and its two nearest and highest-illuminated neighbors, providing three images per event.  The image for the highest-charge string on a candidate signal event is shown in Fig.~\ref{fig:evt_display} (left).  
The three images were then processed by CNNs, trained to distinguish images produced by simulated signal and background events and based on VGG16~\cite{Simonyan:2014cmh}, with a total of $\cal{O}$(100~M) trainable parameters for the high-dimensional signal parameter space. 
Three separate CNNs were used to distinguish the $\nu_\tau$ signal from remaining backgrounds produced by 1) single cascade neutrino interactions such as $\nu_{e,\mu,\tau}$ neutral current (NC) and $\nu_e$ CC, 2) downward-going muons ($\mu_\downarrow$), and 3) both $\nu_\mu$ interactions producing muon tracks and $\mu_\downarrow$; the associated CNN scores are denoted $C_1$, $C_2$ and $C_3$, respectively, with ranges [0,1]. Figure~\ref{fig:evt_display} (right) shows $S(C_1)$, the saliency~\cite{simonyan2014deep} for $C_1$, here defined as the magnitude of the gradient of the CNN score (scaled to [0,1]) of $C_1$ with respect to the signal amplitude at each pixel.  For reference, the contour (solid line) shows where the detected light falls to zero, and is essentially an outline of the plot on the left. (Points outside the contour are variously acausally early, very late, or at distances that are many absorption lengths from the event vertices.)  Large $S(C_1)$ values indicate where and when changes in light level most effectively change $C_1$.  Small $S(C_1)$ values appear in highly-illuminated regions and in regions with no light. Bright regions contribute to $C_1$, but $C_1$ is not as sensitive there to changes in light level as at the leading and trailing edge envelopes of the light from the event, which are roughly coincident with the contour.  The saliency thus shows that $C_1$ is sensitive to the overall shape of emitted light in the detector.

\begin{figure}[h]
\centering
\includegraphics[width=0.95\columnwidth]{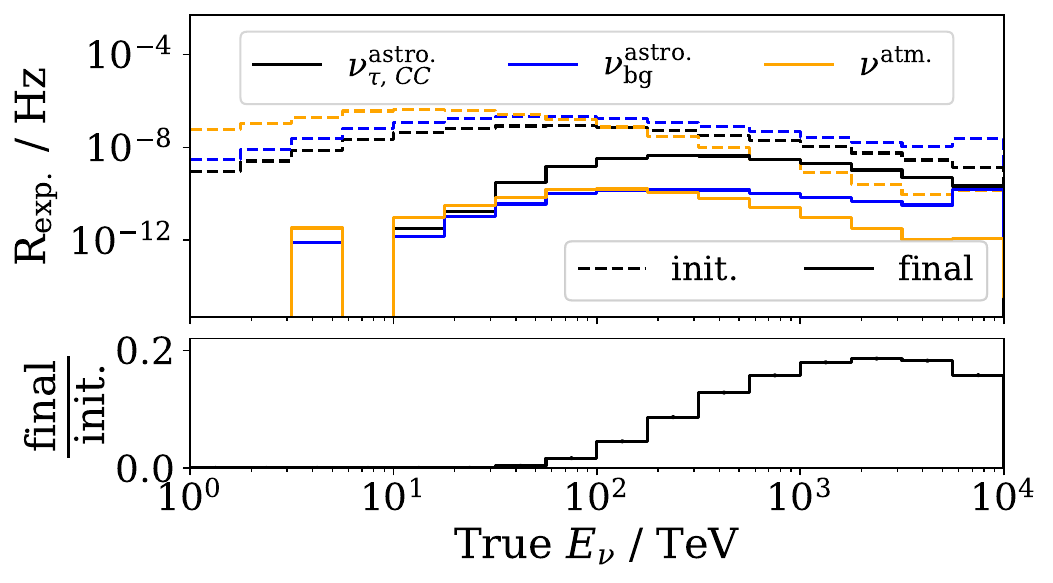}
\caption{Top: Expected rate vs. energy of $\nu_{\tau}^\mathrm{astro}$ CC events, astrophysical and atmospheric neutrino backgrounds with initial selection criteria applied (dashed) and with final selection criteria then also applied (solid); for $\nu^\mathrm{astro}$ the IceCube GlobalFit flux~\protect\cite{IceCube:2015gsk} was assumed.
Bottom: Ratio of $\nu_{\tau}^\mathrm{astro}$ CC rates after final and initial selection criteria.  (Statistical error bars are too small to be visible.  Although not shown in the plot, the backgrounds were simulated up to $E_\nu =100$~PeV.)  
}
\label{fig:energy_without_and_with_CNN_cuts}
\end{figure}

The scores were calculated for each event, and a signal-to-noise ratio of $\sim\!14$ was obtained by requiring events to have high scores ($C_1 \ge 0.99$, $C_2 \ge 0.98$ and $C_3 \ge 0.85$).  The dominant backgrounds come from other $\nu^\mathrm{astro}$ flavors and $\nu^\mathrm{atm}$.  The expected energy spectra for signal and the dominant backgrounds, after application of initial and then final selection criteria (including the high CNN scores), are shown in Fig.~\ref{fig:energy_without_and_with_CNN_cuts}.

A sub-dominant ``edge event'' background was observed from simulated cosmic-ray muons that deposited most of their Cherenkov light on a single string on the outer edge of the detector. 
We required $C_3>0.95$ for edge events, reducing this background by about an order of magnitude at an estimated 15\% signal loss.
Table~\ref{tab:PredictedEvents} lists the expected number of events, after application of the initial and final sets of selection criteria, assuming the best-fit parameters from two IceCube flux measurements.

The largest backgrounds are due to other astrophysical neutrino interactions, and conventional and prompt atmospheric neutrinos, followed by muons from $\pi^\pm/K^\pm$ decays in cosmic-ray air showers. The backgrounds listed in Table~\ref{tab:PredictedEvents} were estimated using simulation packages for astrophysical neutrinos~\cite{anis}, muons from cosmic-ray air showers~\cite{Heck:1998vt,JakobThesis} (with cosmic-ray primary flux given by~\cite{Gaisser:2011klf} and hadronic interaction model by~\cite{SIBYLL21}), conventional atmospheric neutrino flux from $\pi^\pm/K^\pm$ decays~\cite{HKKMS06} following our published $\nu^\mathrm{atm}$ flux measurements above $E_\nu\! \sim\! 50$~TeV~\cite{ICAtm2011, ICAtm2013,ICAtm2015}, and prompt atmospheric neutrino flux~\cite{IceCube:2020wum,BERSS,Garzelli:2015psa,Gauld:2015kvh} postulated to arise from the decays of charm or heavier mesons produced in air-showers and modeled following Ref.~\cite{BERSS}.  Electromagnetic (EM) and hadronic showers below 1~PeV were simulated based on the parameterizations of the mean longitudinal and lateral profiles in Ref.~\cite{Radel:2012ij} and included fluctuations in the energy of the hadronic component. Above 1~PeV the LPM effect~\cite{Landau:1953gr,Landau:1953um,Migdal:1956tc} is used for EM showers. (Our treatment of possible prompt atmospheric muons is described in the Appendix.) The total $\nu$N deep inelastic scattering cross section is from~\cite{sarkar}.

\begin{widetext}
\onecolumngrid
{\renewcommand{\arraystretch}{1.3}
\begin{table}[b]
\small
\centering
\begin{tabular}{|l||c||c|c|c|c|c|}\hline
      & $\nu_{\tau, CC}^\mathrm{astro}$~\cite{anis} 
         & $\nu_\mathrm{other}^\mathrm{astro}$  ~\cite{anis}
         & $\nu_\mathrm{conv.}^\mathrm{atm}$~\cite{HKKMS06,ICAtm2011,ICAtm2013,ICAtm2015}  
         & $\nu_\mathrm{prompt}^\mathrm{atm}$~\cite{IceCube:2020wum,BERSS,Garzelli:2015psa,Gauld:2015kvh}  
         & $\mu_\mathrm{conv.}^\mathrm{atm}$~\cite{Heck:1998vt,JakobThesis,Gaisser:2011klf,SIBYLL21}  
         & all background \\ \hline
 initial & $160 \pm 0.2$ ($190 \pm 0.3$)& $400 \pm 0.7$ ($490 \pm 0.8$) & $580 \pm 7$    & $72 \pm 0.1$ 
 & $8400 \pm 110$    & $ 9450 \pm 110$ ($ 9540 \pm 110$) \\ \hline
 final & $6.4 \pm 0.02$ ($4.0 \pm 0.02$) & $0.3 \pm 0.02$ ($0.2 \pm 0.01$) & $0.1 \pm 0.008$ & $0.1 \pm 0.001$ & $0.01 \pm 0.008$ & $0.5 \pm 0.02$ ($0.4 \pm 0.02$) \\ \hline
\end{tabular}
\caption{Expected number of events after initial and final set of selection criteria (including all corrections described in the text) for signal ($\nu_{\tau,\,\mathrm{CC}}^\mathrm{astro}$) and backgrounds, assuming IceCube's flux from Refs.~\cite{IceCube:2015gsk} and (in parentheses)~\cite{IceCube:2020wum}. About 85\% of the estimated contribution from $\nu_\mathrm{prompt}^\mathrm{atm}$ is from $\nu_\tau$.
Signal and astrophysical background levels vary with the flux.  The simulation did not include the self-veto effect~\cite{Schonert:2008is} that would reduce the conventional (conv.) and prompt $\nu^\mathrm{atm}$ backgrounds.  References to associated simulation packages are given; see text for details.  Errors are statistical only, arising from finite simulation samples.}
\label{tab:PredictedEvents}
\end{table}
}
\end{widetext}
\twocolumngrid

Additional potential background from muon deep inelastic scattering ($\mu$ DIS), given by $\mu + X \rightarrow \nu_\mu + X^\prime$, where the light from the incoming $\mu$ followed by the light from the hadronic cascade could mimic the $\nu_\tau$ signature, is estimated from the predicted atmospheric $\nu_\mu$ CC background.  At energies above roughly 100~TeV, we expect comparable numbers of atmospheric $\nu_\mu$ and $\mu$~\cite{Schonert:2008is}, but the $\mu$ energies will be diminished as they pass through the ice to the detector, decreasing their ability to mimic the $\nu_\tau$ signature.  We conservatively doubled the estimated background from atmospheric $\nu_\mu$ CC interactions, from 0.005 to 0.01, to account for the potential background.

We also estimated the background expected from charmed hadrons produced in energetic $\nu_e$ CC and $\nu$ NC interactions.
This background component had not initially been considered in designing the analysis. After unblinding, we became aware of recent results~\cite{ATLAS:2021vod} that indicate that the strange sea in the nucleon is not as suppressed as had been previously believed, so that charm production would thereby be somewhat enhanced compared to our original estimate.
Using a simulated neutrino dataset based on the HERAPDF1.5~\cite{Radescu:2010zz} parton distribution functions (PDFs), and applying a modest correction to reflect more modern PDFs~\cite{Hou:2019efy,Bailey:2020ooq,Faura:2020oom,ATLAS:2016nqi,ATLAS:2021vod}, the estimated background from $\nu^\mathrm{astro}$ increases by 23\% relative to the simulations excluding these interactions.
The theoretical uncertainty from the PDFs at the 100~TeV scale is roughly 3\%, so the increase corresponds to only about $(15 \pm 0.5)\%$ ($0.08 \pm 0.002$ events) of the total background estimation.
We included this additional background directly to maintain our blindness protocol that disallowed retraining the CNNs to reject charm background.
Uncertainties in the cross section for the interaction of charmed mesons and baryons with ordinary matter had a negligible impact. Backgrounds from on-shell $W$ production~\cite{Zhou:2019frk} from high-energy $\nu_e/\nu_\mu$ interactions, top-quark decay and Glashow resonance interactions~\cite{Soto:2021vdc} can produce energetic $\nu_\tau$ or $\tau$, but are collectively estimated to contribute roughly an order of magnitude fewer background events than other sources and were not included in our background estimate.

For the range of astrophysical neutrino fluxes measured by IceCube
(denoted $\phi_\mathrm{astro}^\mathrm{IC}$), and for a 1:1:1 neutrino flavor ratio at the detector, we predicted a final sample of 4--8 $\nu_{\tau}$ CC signal events.  Similarly, the predicted total background varied for each $\phi_\mathrm{astro}^\mathrm{IC}$. Using IceCube's previous measurements of the spectral index $\gamma_\mathrm{astro}$, this relatively small number of events constrains the $\nu^\mathrm{astro}$ flux normalization $\phi_\mathrm{astro}$.  Data satisfying $C_2 > 0.98$ were placed in $4 \!\times\! 4$ bins in their $C_1$ and $C_3$ scores. 

We calculated confidence intervals following Ref.~\cite{Feldman:1997qc} and using the test statistic defined as $\mathrm{TS}(\lambda_\tau) = \ln{L(\hat{\lambda}_\tau}) - \ln{L(\lambda_\tau})$, where $\lambda_\tau = \phi(\nu_{\tau}^\mathrm{astro})/\phi_\mathrm{nom.}(\nu_{\tau}^\mathrm{astro})$, the measured-to-nominal flux ratio. Here the nominal flux is one of the four IceCube measured values, and $\hat{\lambda}_\tau$ the value of $\lambda_\tau$ that maximizes the Poisson likelihood $L$ across all 16 bins.  Critical values were extracted at the desired confidence level using the TS distributions from a range of $\lambda_\tau$ values, each of which were simulated with $10^4$ pseudo experiments.  This procedure incorporated as nuisance parameters the systematic uncertainties in the estimated fluxes for each background component (prior width of 30\% for $\nu^\mathrm{atm}$ and $\nu^\mathrm{astro}$; 50\% for cosmic-ray muons), the detection efficiency of the DOMs (10\%), and the optical scattering properties of the ice (5\%).  Since many of these parameters are degenerate in their effect on the analysis observables, and we expected fewer signal events than nuisance parameters, we estimated their impact by incorporating randomized versions of the parameters for each of the pseudo experiments used to calculate the critical value of our TS.  This procedure increased the critical values relative to their values in the absence of the systematic uncertainties, widening the extracted confidence intervals.

\begin{figure}[h]
\centering
\includegraphics[width=0.48\textwidth]{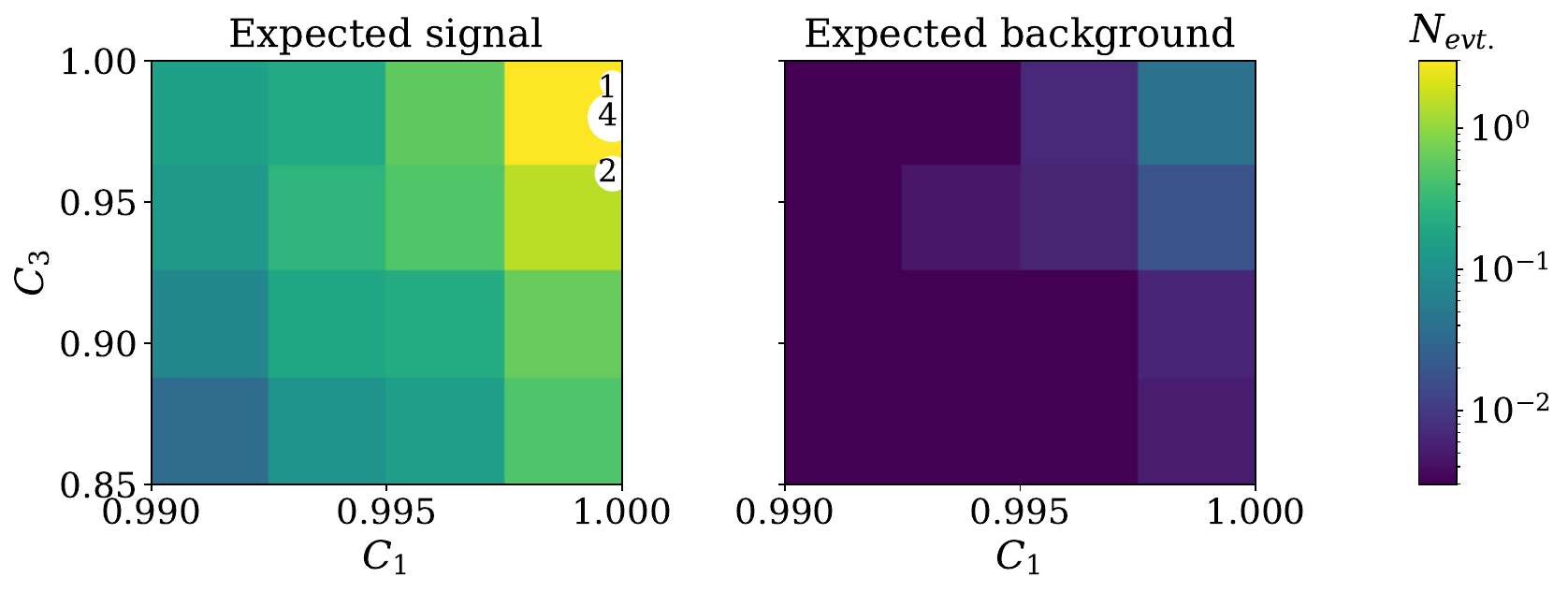}
\caption{Histogram of the $C_3$ vs. $C_1$ CNN scores with all selection criteria applied. The color in each bin gives the expected number of signal (left) and background (right) events in that bin, assuming IceCube's GlobalFit flux~\protect\cite{IceCube:2015gsk}.  The approximate ($C_1$,$C_3$) values of the seven observed candidate $\nu_{\tau}^\mathrm{astro}$ are shown by white circles, with the number inside each circle indicating the number of candidate events there.}
\label{fig:net3_vs_net1}
\end{figure}

Seven events remained after applying the final set of selection criteria to the data, consistent with expectation.  
Figure~\ref{fig:net3_vs_net1} shows the final expected signal and background, assuming IceCube's GlobalFit flux, as a function of $C_3$ vs. $C_1$. Five of the candidate $\nu_\tau$ events are in the upper right bin and two are in the bin just below it.
Three of the seven events were seen in previous IceCube analyses~\cite{IceCube:2013cdw,Lu:2017nti,IceCube:2020fpi,Meier:2019ypu}, and one of these three had previously been identified~\cite{IceCube:2020fpi,Meier:2019ypu} as a candidate $\nu_{\tau}^\mathrm{astro}$.  
For each candidate event we evaluated the ``tauness'' as $P_\tau(i) = n_s(i)/(n_s(i) + n_b(i))$, where $n_s$ and $n_b$ are the expected signal and background in bin $i$ (see Fig.~\ref{fig:net3_vs_net1}). $P_\tau$ ranges from $0.90-0.92$ for two of the candidate $\nu_\tau$, and $0.94-0.95$ for the other five, depending on the assumed $\phi_\mathrm{astro}^\mathrm{IC}$.
For IceCube's GlobalFit flux we predict a total background of $0.5 \pm 0.02$ events (see Table~\ref{tab:PredictedEvents}); using the distribution of the seven observed events and expected backgrounds in the 16 bins in Fig.~\ref{fig:net3_vs_net1}, we exclude the null hypothesis of no $\nu_\tau^\mathrm{astro}$ at a (single-sided) significance of 5.1$\sigma$.  Under the other three flux assumptions~\cite{IceCube:2021uhz,IceCube:2018pgc,IceCube:2020wum}, the significances are $5.2\sigma$, $5.2\sigma$ and $5.5\sigma$, respectively.
The best-fit $\nu_\tau$ flux normalizations are all within the 68\% frequentist confidence intervals of the four IceCube fluxes.

We performed multiple checks on the candidate events to ensure they were consistent with expectation.  For simplicity and to avoid introducing additional systematic uncertainties, the analysis did not employ a tailored $\nu_\tau^\mathrm{astro}$ reconstruction.  However, as a post-unblinding check we used a reconstruction for single-cascade events~\cite{IceCube:2021umt} to estimate the 
energies and directions (Fig.~\ref{fig:EReco_coszenReco}) and vertex positions (see the Appendix).
\begin{figure}[h]
\centering
\begin{minipage}[b]{0.90\columnwidth}
    \includegraphics[width=\linewidth]{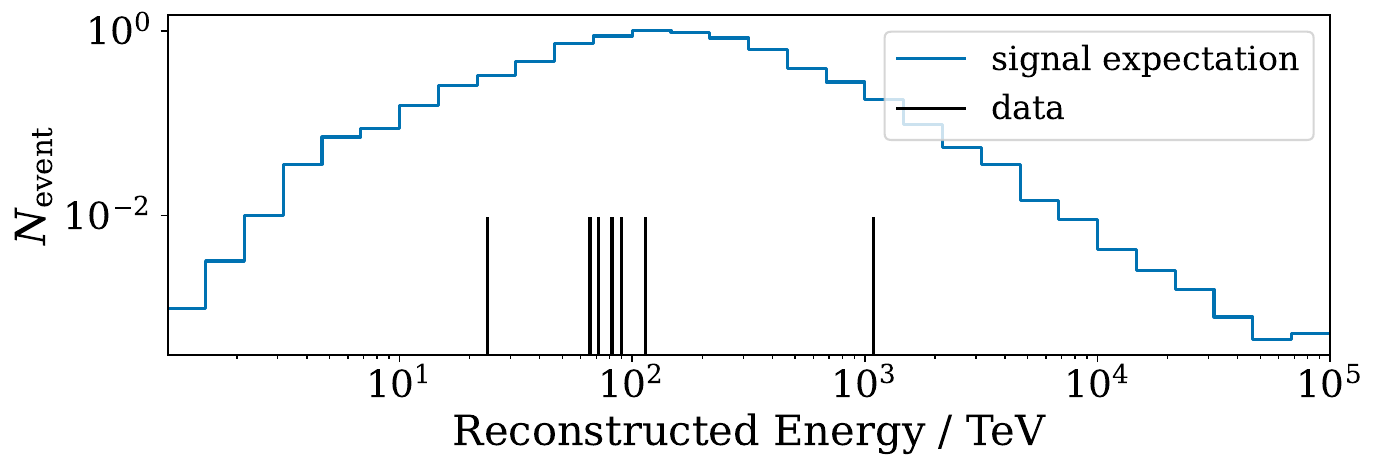}
    \label{fig:EReco}
\end{minipage}
\begin{minipage}[b]{0.90\columnwidth}
    \includegraphics[width=\linewidth]{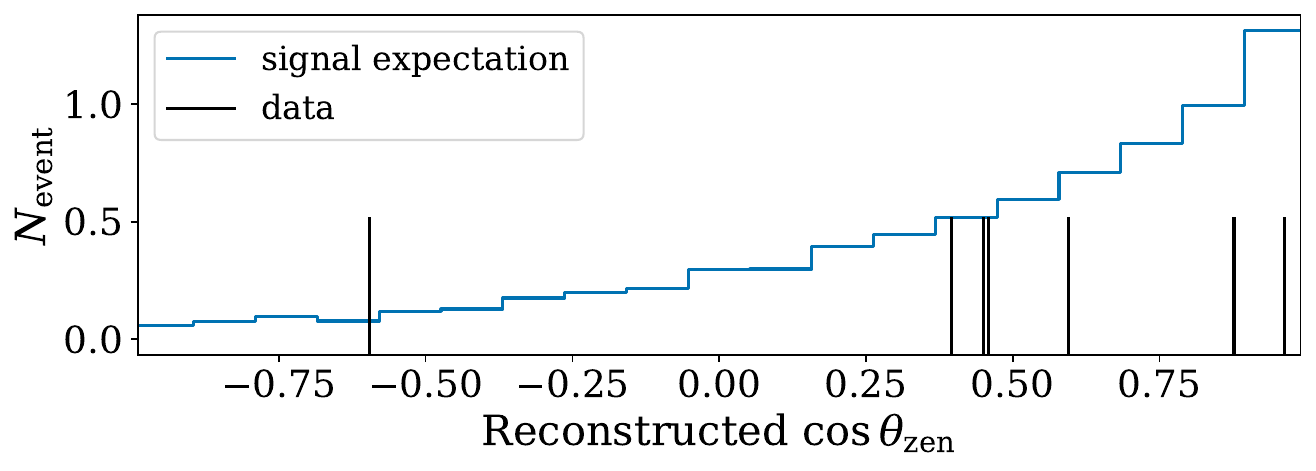}
    \label{fig:coszenReco}
\end{minipage}
\vspace{-0.5cm}
\caption{Reconstructed visible energies (top) and $\cos{\theta_\mathrm{zen}}$ (bottom) for simulated $\nu_\tau^\mathrm{CC}$ (solid histogram) and seven candidate events (vertical lines) for the flux in Ref.~\protect\cite{IceCube:2015gsk}.
The upward-going event with $\cos{\theta_\mathrm{zen}} \simeq -0.6$ had a reconstructed energy of $\sim\! 90$~TeV.}
\label{fig:EReco_coszenReco}
\end{figure}
The median expected $E_{\nu_\tau}$ was roughly 200~TeV (for the flux in Ref.~\cite{IceCube:2020wum}).
The dominant up-down asymmetry is due to Earth absorption and consistent with expectation.  Other polar angle effects such as higher vertical vs. horizontal DOM density and $\nu_\tau$ regeneration~\cite{Halzen:1998be} are also included.
(Simulations predict that for $E_{\tau} \lesssim 0.5$~PeV, the selected events are biased toward higher average $\tau$ decay lengths $\left\langle L_\tau \right\rangle$; \emph{e.g.,} for $E_\tau \sim 100$~TeV, $\left\langle L_\tau \right\rangle \sim 10$~m.)  The events were more clustered in depth than expected but were consistent with a statistical fluctuation, as discussed in the Appendix.
We observed no significant coincident activity in the IceTop cosmic-ray air-shower surface array for any of the events.

We tested the robustness of the CNNs to hypothetical improperly modeled uncertainties by evaluating their susceptibility to correlated and uncorrelated variations of the raw data underlying the images.  
We found that the probability of background-to-signal migration was $< (2 \pm 0.2)\cdot 10^{-5}$ and of signal-to-background migration $< (3 \pm 0.8) \cdot 10^{-3}$. We also employed targeted tests to estimate the CNNs' robustness against less likely changes in the underlying raw data, including adversarial attacks~\cite{moosavi2016deepfool} against candidate signal and simulated background events.  We found that events only migrated in response to changes outside our uncertainty envelope.  These tests are described in the Appendix.   We conclude that the CNNs are robust against detector systematic effects that could present as either uncorrelated or correlated changes in light levels in one or more DOMs, or entire strings, in the detector.

Energetic astrophysical sources, in conjunction with neutrino oscillations over cosmic baselines, provide the only known way to produce large numbers of $\nu_\tau$ energetic enough to create the observed event morphologies. The result presented in this Letter demonstrates that astrophysical $\nu_\tau$ consistent with this hypothesis are present in the IceCube data and provides powerful confirmation of the earlier IceCube discovery of astrophysical neutrinos~\cite{ehe-prl-2013,HESE2y,HESE3y}.  

\newpage
\begin{acknowledgements}
The IceCube Collaboration acknowledges the significant contributions to this manuscript from the Pennsylvania State University.  We dedicate this paper to the memory of Lovisa Arnesson-Cronhamre, a young graduate student whose tragic and untimely passing stole from us all a promising neutrino physicist.
USA {\textendash} U.S. National Science Foundation-Office of Polar Programs,
U.S. National Science Foundation-Physics Division,
U.S. National Science Foundation-EPSCoR, U.S. National Science Foundation-Major Research Instrumentation Program,
Deep Learning for Statistics, Astrophysics, Geoscience, Engineering, Meteorology and Atmospheric Science, Physical Sciences and Psychology (DL-SAGEMAPP) at the Institute for Computational and Data Sciences (ICDS) at the Pennsylvania State University,
Wisconsin Alumni Research Foundation,
Center for High Throughput Computing (CHTC) at the University of Wisconsin{\textendash}Madison,
Open Science Grid (OSG),
Advanced Cyberinfrastructure Coordination Ecosystem: Services {\&} Support (ACCESS),
Frontera computing project at the Texas Advanced Computing Center,
U.S. Department of Energy-National Energy Research Scientific Computing Center,
Particle astrophysics research computing center at the University of Maryland,
Institute for Cyber-Enabled Research at Michigan State University,
and Astroparticle physics computational facility at Marquette University;
Belgium {\textendash} Funds for Scientific Research (FRS-FNRS and FWO),
FWO Odysseus and Big Science programmes,
and Belgian Federal Science Policy Office (Belspo);
Germany {\textendash} Bundesministerium f{\"u}r Bildung und Forschung (BMBF),
Deutsche Forschungsgemeinschaft (DFG),
Helmholtz Alliance for Astroparticle Physics (HAP),
Initiative and Networking Fund of the Helmholtz Association,
Deutsches Elektronen Synchrotron (DESY),
and High Performance Computing cluster of the RWTH Aachen;
Sweden {\textendash} Swedish Research Council,
Swedish Polar Research Secretariat,
Swedish National Infrastructure for Computing (SNIC),
and Knut and Alice Wallenberg Foundation;
European Union {\textendash} EGI Advanced Computing for research;
Australia {\textendash} Australian Research Council;
Canada {\textendash} Natural Sciences and Engineering Research Council of Canada,
Calcul Qu{\'e}bec, Compute Ontario, Canada Foundation for Innovation, WestGrid, and Compute Canada;
Denmark {\textendash} Villum Fonden, Carlsberg Foundation, and European Commission;
New Zealand {\textendash} Marsden Fund;
Japan {\textendash} Japan Society for Promotion of Science (JSPS)
and Institute for Global Prominent Research (IGPR) of Chiba University;
Korea {\textendash} National Research Foundation of Korea (NRF);
Switzerland {\textendash} Swiss National Science Foundation (SNSF);
United Kingdom {\textendash} Department of Physics, University of Oxford.
\end{acknowledgements}


\begin{thebibliography}{100}%
\makeatletter
\providecommand \@ifxundefined [1]{%
 \@ifx{#1\undefined}
}%
\providecommand \@ifnum [1]{%
 \ifnum #1\expandafter \@firstoftwo
 \else \expandafter \@secondoftwo
 \fi
}%
\providecommand \@ifx [1]{%
 \ifx #1\expandafter \@firstoftwo
 \else \expandafter \@secondoftwo
 \fi
}%
\providecommand \natexlab [1]{#1}%
\providecommand \enquote  [1]{``#1''}%
\providecommand \bibnamefont  [1]{#1}%
\providecommand \bibfnamefont [1]{#1}%
\providecommand \citenamefont [1]{#1}%
\providecommand \href@noop [0]{\@secondoftwo}%
\providecommand \href [0]{\begingroup \@sanitize@url \@href}%
\providecommand \@href[1]{\@@startlink{#1}\@@href}%
\providecommand \@@href[1]{\endgroup#1\@@endlink}%
\providecommand \@sanitize@url [0]{\catcode `\\12\catcode `\$12\catcode `\&12\catcode `\#12\catcode `\^12\catcode `\_12\catcode `\%12\relax}%
\providecommand \@@startlink[1]{}%
\providecommand \@@endlink[0]{}%
\providecommand \url  [0]{\begingroup\@sanitize@url \@url }%
\providecommand \@url [1]{\endgroup\@href {#1}{\urlprefix }}%
\providecommand \urlprefix  [0]{URL }%
\providecommand \Eprint [0]{\href }%
\providecommand \doibase [0]{https://doi.org/}%
\providecommand \selectlanguage [0]{\@gobble}%
\providecommand \bibinfo  [0]{\@secondoftwo}%
\providecommand \bibfield  [0]{\@secondoftwo}%
\providecommand \translation [1]{[#1]}%
\providecommand \BibitemOpen [0]{}%
\providecommand \bibitemStop [0]{}%
\providecommand \bibitemNoStop [0]{.\EOS\space}%
\providecommand \EOS [0]{\spacefactor3000\relax}%
\providecommand \BibitemShut  [1]{\csname bibitem#1\endcsname}%
\let\auto@bib@innerbib\@empty
\bibitem [{\citenamefont {Aartsen}\ \emph {et~al.}(2013{\natexlab{a}})\citenamefont {Aartsen} \emph {et~al.}}]{IceCube:2013cdw}%
  \BibitemOpen
  \bibfield  {author} {\bibinfo {author} {\bibfnamefont {M.~G.}\ \bibnamefont {Aartsen}} \emph {et~al.} (\bibinfo {collaboration} {IceCube}),\ }\href {https://doi.org/10.1103/PhysRevLett.111.021103} {\bibfield  {journal} {\bibinfo  {journal} {Phys. Rev. Lett.}\ }\textbf {\bibinfo {volume} {111}},\ \bibinfo {pages} {021103} (\bibinfo {year} {2013}{\natexlab{a}})}\BibitemShut {NoStop}%
\bibitem [{\citenamefont {{IceCube Collaboration}}\ \emph {et~al.}(2013{\natexlab{a}})\citenamefont {{IceCube Collaboration}}, \citenamefont {Aartsen} \emph {et~al.}}]{HESE2y}%
  \BibitemOpen
  \bibfield  {author} {\bibinfo {author} {\bibnamefont {{IceCube Collaboration}}}, \bibinfo {author} {\bibfnamefont {M.~G.}\ \bibnamefont {Aartsen}}, \emph {et~al.},\ }\href@noop {} {\bibfield  {journal} {\bibinfo  {journal} {Science}\ }\textbf {\bibinfo {volume} {342}},\ \bibinfo {pages} {1242856} (\bibinfo {year} {2013}{\natexlab{a}})}\BibitemShut {NoStop}%
\bibitem [{\citenamefont {{IceCube Collaboration}}\ \emph {et~al.}(2014)\citenamefont {{IceCube Collaboration}}, \citenamefont {Aartsen} \emph {et~al.}}]{HESE3y}%
  \BibitemOpen
  \bibfield  {author} {\bibinfo {author} {\bibnamefont {{IceCube Collaboration}}}, \bibinfo {author} {\bibfnamefont {M.~G.}\ \bibnamefont {Aartsen}}, \emph {et~al.},\ }\href@noop {} {\bibfield  {journal} {\bibinfo  {journal} {Phys. Rev. Lett.}\ }\textbf {\bibinfo {volume} {113}},\ \bibinfo {pages} {101101} (\bibinfo {year} {2014})}\BibitemShut {NoStop}%
\bibitem [{\citenamefont {Fermi}(1949)}]{Fermi:1949ee}%
  \BibitemOpen
  \bibfield  {author} {\bibinfo {author} {\bibfnamefont {E.}~\bibnamefont {Fermi}},\ }\href {https://doi.org/10.1103/PhysRev.75.1169} {\bibfield  {journal} {\bibinfo  {journal} {Phys. Rev.}\ }\textbf {\bibinfo {volume} {75}},\ \bibinfo {pages} {1169} (\bibinfo {year} {1949})}\BibitemShut {NoStop}%
\bibitem [{\citenamefont {Bell}(1978)}]{Bell-1978}%
  \BibitemOpen
  \bibfield  {author} {\bibinfo {author} {\bibfnamefont {A.~R.}\ \bibnamefont {Bell}},\ }\href@noop {} {\bibfield  {journal} {\bibinfo  {journal} {MNRAS}\ }\textbf {\bibinfo {volume} {182}},\ \bibinfo {pages} {147} (\bibinfo {year} {1978})}\BibitemShut {NoStop}%
\bibitem [{\citenamefont {Gaisser}(1990)}]{Gaisser1990}%
  \BibitemOpen
  \bibfield  {author} {\bibinfo {author} {\bibfnamefont {T.~K.}\ \bibnamefont {Gaisser}},\ }\href@noop {} {\emph {\bibinfo {title} {Cosmic Rays and Particle Physics}}}\ (\bibinfo  {publisher} {Cambridge University Press},\ \bibinfo {year} {1990})\BibitemShut {NoStop}%
\bibitem [{\citenamefont {{Protheroe}}\ and\ \citenamefont {{Kazanas}}(1983)}]{Protheroe}%
  \BibitemOpen
  \bibfield  {author} {\bibinfo {author} {\bibfnamefont {R.~J.}\ \bibnamefont {{Protheroe}}}\ and\ \bibinfo {author} {\bibfnamefont {D.}~\bibnamefont {{Kazanas}}},\ }\href@noop {} {\bibfield  {journal} {\bibinfo  {journal} {Astrophys. J.}\ }\textbf {\bibinfo {volume} {265}},\ \bibinfo {pages} {620} (\bibinfo {year} {1983})}\BibitemShut {NoStop}%
\bibitem [{\citenamefont {{Kazanas}}\ and\ \citenamefont {{Ellison}}(1986)}]{KazEll}%
  \BibitemOpen
  \bibfield  {author} {\bibinfo {author} {\bibfnamefont {D.}~\bibnamefont {{Kazanas}}}\ and\ \bibinfo {author} {\bibfnamefont {D.~C.}\ \bibnamefont {{Ellison}}},\ }\href {https://doi.org/10.1086/164152} {\bibfield  {journal} {\bibinfo  {journal} {Astrophys. J.}\ }\textbf {\bibinfo {volume} {304}},\ \bibinfo {pages} {178} (\bibinfo {year} {1986})}\BibitemShut {NoStop}%
\bibitem [{\citenamefont {{Sikora}}\ \emph {et~al.}(1987)\citenamefont {{Sikora}} \emph {et~al.}}]{Begelman}%
  \BibitemOpen
  \bibfield  {author} {\bibinfo {author} {\bibfnamefont {M.}~\bibnamefont {{Sikora}}} \emph {et~al.},\ }\href@noop {} {\bibfield  {journal} {\bibinfo  {journal} {Astrophys. J. Lett.}\ }\textbf {\bibinfo {volume} {320}},\ \bibinfo {pages} {L81} (\bibinfo {year} {1987})}\BibitemShut {NoStop}%
\bibitem [{\citenamefont {{Stecker}}\ \emph {et~al.}(1991)\citenamefont {{Stecker}}, \citenamefont {{Done}}, \citenamefont {{Salamon}},\ and\ \citenamefont {{Sommers}}}]{Stecketal}%
  \BibitemOpen
  \bibfield  {author} {\bibinfo {author} {\bibfnamefont {F.~W.}\ \bibnamefont {{Stecker}}}, \bibinfo {author} {\bibfnamefont {C.}~\bibnamefont {{Done}}}, \bibinfo {author} {\bibfnamefont {M.~H.}\ \bibnamefont {{Salamon}}},\ and\ \bibinfo {author} {\bibfnamefont {P.}~\bibnamefont {{Sommers}}},\ }\href@noop {} {\bibfield  {journal} {\bibinfo  {journal} {Phys. Rev. Lett.}\ }\textbf {\bibinfo {volume} {66}},\ \bibinfo {pages} {2697} (\bibinfo {year} {1991})}\BibitemShut {NoStop}%
\bibitem [{\citenamefont {{Stecker}}\ \emph {et~al.}(1992)\citenamefont {{Stecker}}, \citenamefont {{Done}}, \citenamefont {{Salamon}},\ and\ \citenamefont {{Sommers}}}]{StecketalE}%
  \BibitemOpen
  \bibfield  {author} {\bibinfo {author} {\bibfnamefont {F.~W.}\ \bibnamefont {{Stecker}}}, \bibinfo {author} {\bibfnamefont {C.}~\bibnamefont {{Done}}}, \bibinfo {author} {\bibfnamefont {M.~H.}\ \bibnamefont {{Salamon}}},\ and\ \bibinfo {author} {\bibfnamefont {P.}~\bibnamefont {{Sommers}}},\ }\href@noop {} {\bibfield  {journal} {\bibinfo  {journal} {Phys. Rev. Lett.}\ }\textbf {\bibinfo {volume} {69}},\ \bibinfo {pages} {2738(E)} (\bibinfo {year} {1992})}\BibitemShut {NoStop}%
\bibitem [{\citenamefont {{Mannheim}}\ and\ \citenamefont {{Biermann}}(1992)}]{MannBier}%
  \BibitemOpen
  \bibfield  {author} {\bibinfo {author} {\bibfnamefont {K.}~\bibnamefont {{Mannheim}}}\ and\ \bibinfo {author} {\bibfnamefont {P.~L.}\ \bibnamefont {{Biermann}}},\ }\href@noop {} {\bibfield  {journal} {\bibinfo  {journal} {Astron. Astrophys.}\ }\textbf {\bibinfo {volume} {253}},\ \bibinfo {pages} {L21} (\bibinfo {year} {1992})}\BibitemShut {NoStop}%
\bibitem [{\citenamefont {Matthews}\ \emph {et~al.}(2020)\citenamefont {Matthews}, \citenamefont {Bell},\ and\ \citenamefont {Blundell}}]{Matthews:2020lig}%
  \BibitemOpen
  \bibfield  {author} {\bibinfo {author} {\bibfnamefont {J.}~\bibnamefont {Matthews}}, \bibinfo {author} {\bibfnamefont {A.}~\bibnamefont {Bell}},\ and\ \bibinfo {author} {\bibfnamefont {K.}~\bibnamefont {Blundell}},\ }\href {https://doi.org/10.1016/j.newar.2020.101543} {\bibfield  {journal} {\bibinfo  {journal} {New Astron. Rev.}\ }\textbf {\bibinfo {volume} {89}},\ \bibinfo {pages} {101543} (\bibinfo {year} {2020})}\BibitemShut {NoStop}%
\bibitem [{\citenamefont {{Winter}}(2013)}]{Winter2013}%
  \BibitemOpen
  \bibfield  {author} {\bibinfo {author} {\bibfnamefont {W.}~\bibnamefont {{Winter}}},\ }\href@noop {} {\bibfield  {journal} {\bibinfo  {journal} {Phys. Rev.}\ }\textbf {\bibinfo {volume} {D88}},\ \bibinfo {pages} {083007} (\bibinfo {year} {2013})}\BibitemShut {NoStop}%
\bibitem [{\citenamefont {{Murase}}\ \emph {et~al.}(2013)\citenamefont {{Murase}}, \citenamefont {{Ahlers}},\ and\ \citenamefont {{Lacki}}}]{murase}%
  \BibitemOpen
  \bibfield  {author} {\bibinfo {author} {\bibfnamefont {K.}~\bibnamefont {{Murase}}}, \bibinfo {author} {\bibfnamefont {M.}~\bibnamefont {{Ahlers}}},\ and\ \bibinfo {author} {\bibfnamefont {B.~C.}\ \bibnamefont {{Lacki}}},\ }\href@noop {} {\bibfield  {journal} {\bibinfo  {journal} {Phys. Rev.}\ }\textbf {\bibinfo {volume} {D88}},\ \bibinfo {pages} {121301(R)} (\bibinfo {year} {2013})}\BibitemShut {NoStop}%
\bibitem [{\citenamefont {{Learned}}\ and\ \citenamefont {{Pakvasa}}(1995)}]{LearnedPakvasa}%
  \BibitemOpen
  \bibfield  {author} {\bibinfo {author} {\bibfnamefont {J.~G.}\ \bibnamefont {{Learned}}}\ and\ \bibinfo {author} {\bibfnamefont {S.}~\bibnamefont {{Pakvasa}}},\ }\href@noop {} {\bibfield  {journal} {\bibinfo  {journal} {Astropart. Phys.}\ }\textbf {\bibinfo {volume} {3}},\ \bibinfo {pages} {267} (\bibinfo {year} {1995})}\BibitemShut {NoStop}%
\bibitem [{\citenamefont {{Athar}}\ \emph {et~al.}(2006)\citenamefont {{Athar}}, \citenamefont {{Kim}},\ and\ \citenamefont {{Lee}}}]{Athar2006}%
  \BibitemOpen
  \bibfield  {author} {\bibinfo {author} {\bibfnamefont {H.}~\bibnamefont {{Athar}}}, \bibinfo {author} {\bibfnamefont {C.~S.}\ \bibnamefont {{Kim}}},\ and\ \bibinfo {author} {\bibfnamefont {J.}~\bibnamefont {{Lee}}},\ }\href@noop {} {\bibfield  {journal} {\bibinfo  {journal} {Mod. Phys. Lett.}\ }\textbf {\bibinfo {volume} {A21}},\ \bibinfo {pages} {1049} (\bibinfo {year} {2006})}\BibitemShut {NoStop}%
\bibitem [{\citenamefont {{Kashti}}\ and\ \citenamefont {{Waxman}}(2005)}]{KashtiWaxman}%
  \BibitemOpen
  \bibfield  {author} {\bibinfo {author} {\bibfnamefont {T.}~\bibnamefont {{Kashti}}}\ and\ \bibinfo {author} {\bibfnamefont {E.}~\bibnamefont {{Waxman}}},\ }\href@noop {} {\bibfield  {journal} {\bibinfo  {journal} {Phys. Rev. Lett.}\ }\textbf {\bibinfo {volume} {95}},\ \bibinfo {pages} {181101} (\bibinfo {year} {2005})}\BibitemShut {NoStop}%
\bibitem [{\citenamefont {{Klein}}\ \emph {et~al.}(2013)\citenamefont {{Klein}}, \citenamefont {{Mikkelsen}},\ and\ \citenamefont {{Becker Tjus}}}]{Klein2013}%
  \BibitemOpen
  \bibfield  {author} {\bibinfo {author} {\bibfnamefont {S.~R.}\ \bibnamefont {{Klein}}}, \bibinfo {author} {\bibfnamefont {R.~E.}\ \bibnamefont {{Mikkelsen}}},\ and\ \bibinfo {author} {\bibfnamefont {J.}~\bibnamefont {{Becker Tjus}}},\ }\href@noop {} {\bibfield  {journal} {\bibinfo  {journal} {Astrophys. J.}\ }\textbf {\bibinfo {volume} {779}},\ \bibinfo {pages} {106} (\bibinfo {year} {2013})}\BibitemShut {NoStop}%
\bibitem [{\citenamefont {{Lipari}}\ \emph {et~al.}(2007)\citenamefont {{Lipari}}, \citenamefont {{Lusignoli}},\ and\ \citenamefont {{Meloni}}}]{Lipari2007}%
  \BibitemOpen
  \bibfield  {author} {\bibinfo {author} {\bibfnamefont {P.}~\bibnamefont {{Lipari}}}, \bibinfo {author} {\bibfnamefont {M.}~\bibnamefont {{Lusignoli}}},\ and\ \bibinfo {author} {\bibfnamefont {D.}~\bibnamefont {{Meloni}}},\ }\href@noop {} {\bibfield  {journal} {\bibinfo  {journal} {Phys. Rev. D.}\ }\textbf {\bibinfo {volume} {75}},\ \bibinfo {pages} {123005} (\bibinfo {year} {2007})}\BibitemShut {NoStop}%
\bibitem [{\citenamefont {{Bustamante}}\ \emph {et~al.}(2015)\citenamefont {{Bustamante}}, \citenamefont {{Beacom}},\ and\ \citenamefont {{Winter}}}]{Bustamante2015}%
  \BibitemOpen
  \bibfield  {author} {\bibinfo {author} {\bibfnamefont {M.}~\bibnamefont {{Bustamante}}}, \bibinfo {author} {\bibfnamefont {J.~F.}\ \bibnamefont {{Beacom}}},\ and\ \bibinfo {author} {\bibfnamefont {W.}~\bibnamefont {{Winter}}},\ }\href@noop {} {\bibfield  {journal} {\bibinfo  {journal} {Phys. Rev. Lett.}\ }\textbf {\bibinfo {volume} {115}},\ \bibinfo {pages} {161302} (\bibinfo {year} {2015})}\BibitemShut {NoStop}%
\bibitem [{\citenamefont {{Esmaili}}\ and\ \citenamefont {{Farzan}}(2009)}]{Esmaili2009}%
  \BibitemOpen
  \bibfield  {author} {\bibinfo {author} {\bibfnamefont {A.}~\bibnamefont {{Esmaili}}}\ and\ \bibinfo {author} {\bibfnamefont {Y.}~\bibnamefont {{Farzan}}},\ }\href@noop {} {\bibfield  {journal} {\bibinfo  {journal} {Nucl.Phys.}\ }\textbf {\bibinfo {volume} {B821}},\ \bibinfo {pages} {197} (\bibinfo {year} {2009})}\BibitemShut {NoStop}%
\bibitem [{\citenamefont {{Esteban}}\ \emph {et~al.}()\citenamefont {{Esteban}} \emph {et~al.}}]{nufit}%
  \BibitemOpen
  \bibfield  {author} {\bibinfo {author} {\bibfnamefont {I.}~\bibnamefont {{Esteban}}} \emph {et~al.},\ }\href@noop {} {\bibfield  {journal} {\bibinfo  {journal} {JHEP}\ }\textbf {\bibinfo {volume} {01}},\ \bibinfo {pages} {106 (2019)}}\BibitemShut {NoStop}%
\bibitem [{\citenamefont {{NuFIT 4.1}}(2019)}]{nufit2}%
  \BibitemOpen
  \bibfield  {author} {\bibinfo {author} {\bibnamefont {{NuFIT 4.1}}},\ }\href@noop {} {\bibfield  {journal} {\bibinfo  {journal} {www.nu-fit.org}\ } (\bibinfo {year} {2019})}\BibitemShut {NoStop}%
\bibitem [{\citenamefont {Kodama}\ \emph {et~al.}(2008)\citenamefont {Kodama} \emph {et~al.}}]{DONuT:2007bsg}%
  \BibitemOpen
  \bibfield  {author} {\bibinfo {author} {\bibfnamefont {K.}~\bibnamefont {Kodama}} \emph {et~al.} (\bibinfo {collaboration} {DONuT}),\ }\href {https://doi.org/10.1103/PhysRevD.78.052002} {\bibfield  {journal} {\bibinfo  {journal} {Phys. Rev. D}\ }\textbf {\bibinfo {volume} {78}},\ \bibinfo {pages} {052002} (\bibinfo {year} {2008})}\BibitemShut {NoStop}%
\bibitem [{\citenamefont {Agafonova}\ \emph {et~al.}(2019{\natexlab{a}})\citenamefont {Agafonova} \emph {et~al.}}]{OPERA:2019kzo}%
  \BibitemOpen
  \bibfield  {author} {\bibinfo {author} {\bibfnamefont {N.}~\bibnamefont {Agafonova}} \emph {et~al.} (\bibinfo {collaboration} {OPERA}),\ }\href {https://doi.org/10.1103/PhysRevD.100.051301} {\bibfield  {journal} {\bibinfo  {journal} {Phys. Rev. D}\ }\textbf {\bibinfo {volume} {100}},\ \bibinfo {pages} {051301} (\bibinfo {year} {2019}{\natexlab{a}})}\BibitemShut {NoStop}%
\bibitem [{\citenamefont {Li}\ \emph {et~al.}(2018)\citenamefont {Li} \emph {et~al.}}]{Super-Kamiokande:2017edb}%
  \BibitemOpen
  \bibfield  {author} {\bibinfo {author} {\bibfnamefont {Z.}~\bibnamefont {Li}} \emph {et~al.} (\bibinfo {collaboration} {Super-Kamiokande}),\ }\href {https://doi.org/10.1103/PhysRevD.98.052006} {\bibfield  {journal} {\bibinfo  {journal} {Phys. Rev. D}\ }\textbf {\bibinfo {volume} {98}},\ \bibinfo {pages} {052006} (\bibinfo {year} {2018})}\BibitemShut {NoStop}%
\bibitem [{\citenamefont {Aartsen}\ \emph {et~al.}(2019{\natexlab{a}})\citenamefont {Aartsen} \emph {et~al.}}]{IceCube:2019dqi}%
  \BibitemOpen
  \bibfield  {author} {\bibinfo {author} {\bibfnamefont {M.~G.}\ \bibnamefont {Aartsen}} \emph {et~al.} (\bibinfo {collaboration} {IceCube}),\ }\href {https://doi.org/10.1103/PhysRevD.99.032007} {\bibfield  {journal} {\bibinfo  {journal} {Phys. Rev. D}\ }\textbf {\bibinfo {volume} {99}},\ \bibinfo {pages} {032007} (\bibinfo {year} {2019}{\natexlab{a}})}\BibitemShut {NoStop}%
\bibitem [{\citenamefont {Bhattacharya}\ \emph {et~al.}(2016)\citenamefont {Bhattacharya}, \citenamefont {Enberg}, \citenamefont {Jeong}, \citenamefont {Kim}, \citenamefont {Reno}, \citenamefont {Sarcevic},\ and\ \citenamefont {Stasto}}]{Bhattacharya:2016jce}%
  \BibitemOpen
  \bibfield  {author} {\bibinfo {author} {\bibfnamefont {A.}~\bibnamefont {Bhattacharya}}, \bibinfo {author} {\bibfnamefont {R.}~\bibnamefont {Enberg}}, \bibinfo {author} {\bibfnamefont {Y.~S.}\ \bibnamefont {Jeong}}, \bibinfo {author} {\bibfnamefont {C.~S.}\ \bibnamefont {Kim}}, \bibinfo {author} {\bibfnamefont {M.~H.}\ \bibnamefont {Reno}}, \bibinfo {author} {\bibfnamefont {I.}~\bibnamefont {Sarcevic}},\ and\ \bibinfo {author} {\bibfnamefont {A.}~\bibnamefont {Stasto}},\ }\href {https://doi.org/10.1007/JHEP11(2016)167} {\bibfield  {journal} {\bibinfo  {journal} {JHEP}\ }\textbf {\bibinfo {volume} {11}},\ \bibinfo {pages} {16 (2016)}}\BibitemShut {NoStop}%
\bibitem [{\citenamefont {Arg\"uelles}\ \emph {et~al.}(2015)\citenamefont {Arg\"uelles}, \citenamefont {Katori},\ and\ \citenamefont {Salvado}}]{Arguelles:2015dca}%
  \BibitemOpen
  \bibfield  {author} {\bibinfo {author} {\bibfnamefont {C.~A.}\ \bibnamefont {Arg\"uelles}}, \bibinfo {author} {\bibfnamefont {T.}~\bibnamefont {Katori}},\ and\ \bibinfo {author} {\bibfnamefont {J.}~\bibnamefont {Salvado}},\ }\href {https://doi.org/10.1103/PhysRevLett.115.161303} {\bibfield  {journal} {\bibinfo  {journal} {Phys. Rev. Lett.}\ }\textbf {\bibinfo {volume} {115}},\ \bibinfo {pages} {161303} (\bibinfo {year} {2015})}\BibitemShut {NoStop}%
\bibitem [{\citenamefont {Pagliaroli}\ \emph {et~al.}(2015)\citenamefont {Pagliaroli}, \citenamefont {Palladino}, \citenamefont {Villante},\ and\ \citenamefont {Vissani}}]{Pagliaroli:2015rca}%
  \BibitemOpen
  \bibfield  {author} {\bibinfo {author} {\bibfnamefont {G.}~\bibnamefont {Pagliaroli}}, \bibinfo {author} {\bibfnamefont {A.}~\bibnamefont {Palladino}}, \bibinfo {author} {\bibfnamefont {F.~L.}\ \bibnamefont {Villante}},\ and\ \bibinfo {author} {\bibfnamefont {F.}~\bibnamefont {Vissani}},\ }\href {https://doi.org/10.1103/PhysRevD.92.113008} {\bibfield  {journal} {\bibinfo  {journal} {Phys. Rev. D}\ }\textbf {\bibinfo {volume} {92}},\ \bibinfo {pages} {113008} (\bibinfo {year} {2015})}\BibitemShut {NoStop}%
\bibitem [{\citenamefont {Shoemaker}\ and\ \citenamefont {Murase}(2016)}]{Shoemaker:2015qul}%
  \BibitemOpen
  \bibfield  {author} {\bibinfo {author} {\bibfnamefont {I.~M.}\ \bibnamefont {Shoemaker}}\ and\ \bibinfo {author} {\bibfnamefont {K.}~\bibnamefont {Murase}},\ }\href {https://doi.org/10.1103/PhysRevD.93.085004} {\bibfield  {journal} {\bibinfo  {journal} {Phys. Rev. D}\ }\textbf {\bibinfo {volume} {93}},\ \bibinfo {pages} {085004} (\bibinfo {year} {2016})}\BibitemShut {NoStop}%
\bibitem [{\citenamefont {Brdar}\ \emph {et~al.}(2017)\citenamefont {Brdar}, \citenamefont {Kopp},\ and\ \citenamefont {Wang}}]{Brdar:2016thq}%
  \BibitemOpen
  \bibfield  {author} {\bibinfo {author} {\bibfnamefont {V.}~\bibnamefont {Brdar}}, \bibinfo {author} {\bibfnamefont {J.}~\bibnamefont {Kopp}},\ and\ \bibinfo {author} {\bibfnamefont {X.-P.}\ \bibnamefont {Wang}},\ }\href {https://doi.org/10.1088/1475-7516/2017/01/026} {\bibfield  {journal} {\bibinfo  {journal} {JCAP}\ }\textbf {\bibinfo {volume} {01}},\ \bibinfo {pages} {026 (2017)}}\BibitemShut {NoStop}%
\bibitem [{\citenamefont {Bustamante}\ \emph {et~al.}(2017)\citenamefont {Bustamante}, \citenamefont {Beacom},\ and\ \citenamefont {Murase}}]{Bustamante:2016ciw}%
  \BibitemOpen
  \bibfield  {author} {\bibinfo {author} {\bibfnamefont {M.}~\bibnamefont {Bustamante}}, \bibinfo {author} {\bibfnamefont {J.~F.}\ \bibnamefont {Beacom}},\ and\ \bibinfo {author} {\bibfnamefont {K.}~\bibnamefont {Murase}},\ }\href {https://doi.org/10.1103/PhysRevD.95.063013} {\bibfield  {journal} {\bibinfo  {journal} {Phys. Rev. D}\ }\textbf {\bibinfo {volume} {95}},\ \bibinfo {pages} {063013} (\bibinfo {year} {2017})}\BibitemShut {NoStop}%
\bibitem [{\citenamefont {Klop}\ and\ \citenamefont {Ando}(2018)}]{Klop:2017dim}%
  \BibitemOpen
  \bibfield  {author} {\bibinfo {author} {\bibfnamefont {N.}~\bibnamefont {Klop}}\ and\ \bibinfo {author} {\bibfnamefont {S.}~\bibnamefont {Ando}},\ }\href {https://doi.org/10.1103/PhysRevD.97.063006} {\bibfield  {journal} {\bibinfo  {journal} {Phys. Rev. D}\ }\textbf {\bibinfo {volume} {97}},\ \bibinfo {pages} {063006} (\bibinfo {year} {2018})}\BibitemShut {NoStop}%
\bibitem [{\citenamefont {Rasmussen}\ \emph {et~al.}(2017)\citenamefont {Rasmussen}, \citenamefont {Lechner}, \citenamefont {Ackermann}, \citenamefont {Kowalski},\ and\ \citenamefont {Winter}}]{Rasmussen:2017ert}%
  \BibitemOpen
  \bibfield  {author} {\bibinfo {author} {\bibfnamefont {R.~W.}\ \bibnamefont {Rasmussen}}, \bibinfo {author} {\bibfnamefont {L.}~\bibnamefont {Lechner}}, \bibinfo {author} {\bibfnamefont {M.}~\bibnamefont {Ackermann}}, \bibinfo {author} {\bibfnamefont {M.}~\bibnamefont {Kowalski}},\ and\ \bibinfo {author} {\bibfnamefont {W.}~\bibnamefont {Winter}},\ }\href {https://doi.org/10.1103/PhysRevD.96.083018} {\bibfield  {journal} {\bibinfo  {journal} {Phys. Rev. D}\ }\textbf {\bibinfo {volume} {96}},\ \bibinfo {pages} {083018} (\bibinfo {year} {2017})}\BibitemShut {NoStop}%
\bibitem [{\citenamefont {Bustamante}\ and\ \citenamefont {Agarwalla}(2019)}]{Bustamante:2018mzu}%
  \BibitemOpen
  \bibfield  {author} {\bibinfo {author} {\bibfnamefont {M.}~\bibnamefont {Bustamante}}\ and\ \bibinfo {author} {\bibfnamefont {S.~K.}\ \bibnamefont {Agarwalla}},\ }\href {https://doi.org/10.1103/PhysRevLett.122.061103} {\bibfield  {journal} {\bibinfo  {journal} {Phys. Rev. Lett.}\ }\textbf {\bibinfo {volume} {122}},\ \bibinfo {pages} {061103} (\bibinfo {year} {2019})}\BibitemShut {NoStop}%
\bibitem [{\citenamefont {Denton}\ and\ \citenamefont {Tamborra}(2018)}]{Denton:2018aml}%
  \BibitemOpen
  \bibfield  {author} {\bibinfo {author} {\bibfnamefont {P.~B.}\ \bibnamefont {Denton}}\ and\ \bibinfo {author} {\bibfnamefont {I.}~\bibnamefont {Tamborra}},\ }\href {https://doi.org/10.1103/PhysRevLett.121.121802} {\bibfield  {journal} {\bibinfo  {journal} {Phys. Rev. Lett.}\ }\textbf {\bibinfo {volume} {121}},\ \bibinfo {pages} {121802} (\bibinfo {year} {2018})}\BibitemShut {NoStop}%
\bibitem [{\citenamefont {Farzan}\ and\ \citenamefont {Palomares-Ruiz}(2019)}]{Farzan:2018pnk}%
  \BibitemOpen
  \bibfield  {author} {\bibinfo {author} {\bibfnamefont {Y.}~\bibnamefont {Farzan}}\ and\ \bibinfo {author} {\bibfnamefont {S.}~\bibnamefont {Palomares-Ruiz}},\ }\href {https://doi.org/10.1103/PhysRevD.99.051702} {\bibfield  {journal} {\bibinfo  {journal} {Phys. Rev. D}\ }\textbf {\bibinfo {volume} {99}},\ \bibinfo {pages} {051702} (\bibinfo {year} {2019})}\BibitemShut {NoStop}%
\bibitem [{\citenamefont {Abbasi}\ \emph {et~al.}(2022{\natexlab{a}})\citenamefont {Abbasi} \emph {et~al.}}]{IceCube:2021tdn}%
  \BibitemOpen
  \bibfield  {author} {\bibinfo {author} {\bibfnamefont {R.}~\bibnamefont {Abbasi}} \emph {et~al.} (\bibinfo {collaboration} {IceCube}),\ }\href {https://doi.org/10.1038/s41567-022-01762-1} {\bibfield  {journal} {\bibinfo  {journal} {Nature Phys.}\ }\textbf {\bibinfo {volume} {18}},\ \bibinfo {pages} {1287} (\bibinfo {year} {2022}{\natexlab{a}})}\BibitemShut {NoStop}%
\bibitem [{\citenamefont {Abdullahi}\ and\ \citenamefont {Denton}(2020)}]{Abdullahi:2020rge}%
  \BibitemOpen
  \bibfield  {author} {\bibinfo {author} {\bibfnamefont {A.}~\bibnamefont {Abdullahi}}\ and\ \bibinfo {author} {\bibfnamefont {P.~B.}\ \bibnamefont {Denton}},\ }\href {https://doi.org/10.1103/PhysRevD.102.023018} {\bibfield  {journal} {\bibinfo  {journal} {Phys. Rev. D}\ }\textbf {\bibinfo {volume} {102}},\ \bibinfo {pages} {023018} (\bibinfo {year} {2020})}\BibitemShut {NoStop}%
\bibitem [{\citenamefont {Abraham}\ \emph {et~al.}(2022)\citenamefont {Abraham} \emph {et~al.}}]{Abraham:2022jse}%
  \BibitemOpen
  \bibfield  {author} {\bibinfo {author} {\bibfnamefont {R.~M.}\ \bibnamefont {Abraham}} \emph {et~al.},\ }\href {https://doi.org/10.1088/1361-6471/ac89d2} {\bibfield  {journal} {\bibinfo  {journal} {Journal of Physics G: Nuclear and Particle Physics}\ }\textbf {\bibinfo {volume} {49}},\ \bibinfo {pages} {110501} (\bibinfo {year} {2022})}\BibitemShut {NoStop}%
\bibitem [{\citenamefont {Ackermann}\ \emph {et~al.}(2022)\citenamefont {Ackermann} \emph {et~al.}}]{Ackermann:2022rqc}%
  \BibitemOpen
  \bibfield  {author} {\bibinfo {author} {\bibfnamefont {M.}~\bibnamefont {Ackermann}} \emph {et~al.},\ }\href@noop {} {\bibfield  {journal} {\bibinfo  {journal} {arXiv:2203.08096}\ } (\bibinfo {year} {2022})}\BibitemShut {NoStop}%
\bibitem [{\citenamefont {Abbasi}\ \emph {et~al.}(2012{\natexlab{a}})\citenamefont {Abbasi} \emph {et~al.}}]{PhysRevD.86.022005}%
  \BibitemOpen
  \bibfield  {author} {\bibinfo {author} {\bibfnamefont {R.}~\bibnamefont {Abbasi}} \emph {et~al.} (\bibinfo {collaboration} {IceCube Collaboration}),\ }\href {https://doi.org/10.1103/PhysRevD.86.022005} {\bibfield  {journal} {\bibinfo  {journal} {Phys. Rev. D}\ }\textbf {\bibinfo {volume} {86}},\ \bibinfo {pages} {022005} (\bibinfo {year} {2012}{\natexlab{a}})}\BibitemShut {NoStop}%
\bibitem [{\citenamefont {Aartsen}\ \emph {et~al.}(2016{\natexlab{a}})\citenamefont {Aartsen} \emph {et~al.}}]{PhysRevD.93.022001}%
  \BibitemOpen
  \bibfield  {author} {\bibinfo {author} {\bibfnamefont {M.~G.}\ \bibnamefont {Aartsen}} \emph {et~al.} (\bibinfo {collaboration} {IceCube Collaboration}),\ }\href {https://doi.org/10.1103/PhysRevD.93.022001} {\bibfield  {journal} {\bibinfo  {journal} {Phys. Rev. D}\ }\textbf {\bibinfo {volume} {93}},\ \bibinfo {pages} {022001} (\bibinfo {year} {2016}{\natexlab{a}})}\BibitemShut {NoStop}%
\bibitem [{\citenamefont {Abbasi}\ \emph {et~al.}(2022{\natexlab{b}})\citenamefont {Abbasi} \emph {et~al.}}]{IceCube:2020fpi}%
  \BibitemOpen
  \bibfield  {author} {\bibinfo {author} {\bibfnamefont {R.}~\bibnamefont {Abbasi}} \emph {et~al.} (\bibinfo {collaboration} {IceCube}),\ }\href {https://doi.org/10.1140/epjc/s10052-022-10795-y} {\bibfield  {journal} {\bibinfo  {journal} {Eur. Phys. J. C}\ }\textbf {\bibinfo {volume} {82}},\ \bibinfo {pages} {1031} (\bibinfo {year} {2022}{\natexlab{b}})}\BibitemShut {NoStop}%
\bibitem [{\citenamefont {Meier}\ and\ \citenamefont {Soedingrekso}(2020)}]{Meier:2019ypu}%
  \BibitemOpen
  \bibfield  {author} {\bibinfo {author} {\bibfnamefont {M.}~\bibnamefont {Meier}}\ and\ \bibinfo {author} {\bibfnamefont {J.}~\bibnamefont {Soedingrekso}} (\bibinfo {collaboration} {IceCube}),\ }\href {https://doi.org/10.22323/1.358.0960} {\bibfield  {journal} {\bibinfo  {journal} {PoS ICRC2019}\ ,\ \bibinfo {pages} {960}} (\bibinfo {year} {2020})}\BibitemShut {NoStop}%
\bibitem [{\citenamefont {Aartsen}\ \emph {et~al.}(2017)\citenamefont {Aartsen} \emph {et~al.}}]{IceCube:2016zyt}%
  \BibitemOpen
  \bibfield  {author} {\bibinfo {author} {\bibfnamefont {M.~G.}\ \bibnamefont {Aartsen}} \emph {et~al.} (\bibinfo {collaboration} {IceCube}),\ }\href {https://doi.org/10.1088/1748-0221/12/03/P03012} {\bibfield  {journal} {\bibinfo  {journal} {JINST}\ }\textbf {\bibinfo {volume} {12}}\bibinfo  {number} { (03)},\ \bibinfo {pages} {P03012 (2017)}}\BibitemShut {NoStop}%
\bibitem [{\citenamefont {Abbasi}\ \emph {et~al.}(2012{\natexlab{b}})\citenamefont {Abbasi} \emph {et~al.}}]{IceCube:2011ucd}%
  \BibitemOpen
\bibfield  {number} {  }\bibfield  {author} {\bibinfo {author} {\bibfnamefont {R.}~\bibnamefont {Abbasi}} \emph {et~al.} (\bibinfo {collaboration} {IceCube}),\ }\href {https://doi.org/10.1016/j.astropartphys.2012.01.004} {\bibfield  {journal} {\bibinfo  {journal} {Astropart. Phys.}\ }\textbf {\bibinfo {volume} {35}},\ \bibinfo {pages} {615} (\bibinfo {year} {2012}{\natexlab{b}})}\BibitemShut {NoStop}%
\bibitem [{\citenamefont {Cherenkov}(1934)}]{Cherenkov:1934ilx}%
  \BibitemOpen
  \bibfield  {author} {\bibinfo {author} {\bibfnamefont {P.~A.}\ \bibnamefont {Cherenkov}},\ }\href {https://doi.org/10.3367/UFNr.0093.196710n.0385} {\bibfield  {journal} {\bibinfo  {journal} {Dokl. Akad. Nauk SSSR}\ }\textbf {\bibinfo {volume} {2}},\ \bibinfo {pages} {451} (\bibinfo {year} {1934})}\BibitemShut {NoStop}%
\bibitem [{\citenamefont {Glashow}(1960)}]{gr}%
  \BibitemOpen
  \bibfield  {author} {\bibinfo {author} {\bibfnamefont {S.~L.}\ \bibnamefont {Glashow}},\ }\href@noop {} {\bibfield  {journal} {\bibinfo  {journal} {Phys. Rev.}\ }\textbf {\bibinfo {volume} {118}},\ \bibinfo {pages} {1} (\bibinfo {year} {1960})}\BibitemShut {NoStop}%
\bibitem [{\citenamefont {Aartsen}\ \emph {et~al.}(2021)\citenamefont {Aartsen} \emph {et~al.}}]{IceCube:2021rpz}%
  \BibitemOpen
  \bibfield  {author} {\bibinfo {author} {\bibfnamefont {M.~G.}\ \bibnamefont {Aartsen}} \emph {et~al.} (\bibinfo {collaboration} {IceCube}),\ }\href {https://doi.org/10.1038/s41586-021-03256-1} {\bibfield  {journal} {\bibinfo  {journal} {Nature}\ }\textbf {\bibinfo {volume} {591}},\ \bibinfo {pages} {220} (\bibinfo {year} {2021})},\ \bibinfo {note} {[Erratum: Nature 592, E11 (2021)]}\BibitemShut {NoStop}%
\bibitem [{\citenamefont {Aartsen}\ \emph {et~al.}(2015)\citenamefont {Aartsen} \emph {et~al.}}]{IceCube:2015gsk}%
  \BibitemOpen
  \bibfield  {author} {\bibinfo {author} {\bibfnamefont {M.~G.}\ \bibnamefont {Aartsen}} \emph {et~al.} (\bibinfo {collaboration} {IceCube}),\ }\href {https://doi.org/10.1088/0004-637X/809/1/98} {\bibfield  {journal} {\bibinfo  {journal} {Astrophys. J.}\ }\textbf {\bibinfo {volume} {809}},\ \bibinfo {pages} {98} (\bibinfo {year} {2015})}\BibitemShut {NoStop}%
\bibitem [{\citenamefont {Abbasi}\ \emph {et~al.}(2022{\natexlab{c}})\citenamefont {Abbasi} \emph {et~al.}}]{IceCube:2021uhz}%
  \BibitemOpen
  \bibfield  {author} {\bibinfo {author} {\bibfnamefont {R.}~\bibnamefont {Abbasi}} \emph {et~al.} (\bibinfo {collaboration} {IceCube}),\ }\href {https://doi.org/10.3847/1538-4357/ac4d29} {\bibfield  {journal} {\bibinfo  {journal} {Astrophys. J.}\ }\textbf {\bibinfo {volume} {928}},\ \bibinfo {pages} {50} (\bibinfo {year} {2022}{\natexlab{c}})}\BibitemShut {NoStop}%
\bibitem [{\citenamefont {Aartsen}\ \emph {et~al.}(2019{\natexlab{b}})\citenamefont {Aartsen} \emph {et~al.}}]{IceCube:2018pgc}%
  \BibitemOpen
  \bibfield  {author} {\bibinfo {author} {\bibfnamefont {M.~G.}\ \bibnamefont {Aartsen}} \emph {et~al.} (\bibinfo {collaboration} {IceCube}),\ }\href {https://doi.org/10.1103/PhysRevD.99.032004} {\bibfield  {journal} {\bibinfo  {journal} {Phys. Rev. D}\ }\textbf {\bibinfo {volume} {99}},\ \bibinfo {pages} {032004} (\bibinfo {year} {2019}{\natexlab{b}})}\BibitemShut {NoStop}%
\bibitem [{\citenamefont {Abbasi}\ \emph {et~al.}(2021{\natexlab{a}})\citenamefont {Abbasi} \emph {et~al.}}]{IceCube:2020wum}%
  \BibitemOpen
  \bibfield  {author} {\bibinfo {author} {\bibfnamefont {R.}~\bibnamefont {Abbasi}} \emph {et~al.} (\bibinfo {collaboration} {IceCube}),\ }\href {https://doi.org/10.1103/PhysRevD.104.022002} {\bibfield  {journal} {\bibinfo  {journal} {Phys. Rev. D}\ }\textbf {\bibinfo {volume} {104}},\ \bibinfo {pages} {022002} (\bibinfo {year} {2021}{\natexlab{a}})}\BibitemShut {NoStop}%
\bibitem [{\citenamefont {Simonyan}\ and\ \citenamefont {Zisserman}(2014)}]{Simonyan:2014cmh}%
  \BibitemOpen
  \bibfield  {author} {\bibinfo {author} {\bibfnamefont {K.}~\bibnamefont {Simonyan}}\ and\ \bibinfo {author} {\bibfnamefont {A.}~\bibnamefont {Zisserman}},\ }\href@noop {} {\bibfield  {journal} {\bibinfo  {journal} {arXiv:1409.1556}\ } (\bibinfo {year} {2014})}\BibitemShut {NoStop}%
\bibitem [{\citenamefont {Simonyan}\ \emph {et~al.}(2014)\citenamefont {Simonyan}, \citenamefont {Vedaldi},\ and\ \citenamefont {Zisserman}}]{simonyan2014deep}%
  \BibitemOpen
  \bibfield  {author} {\bibinfo {author} {\bibfnamefont {K.}~\bibnamefont {Simonyan}}, \bibinfo {author} {\bibfnamefont {A.}~\bibnamefont {Vedaldi}},\ and\ \bibinfo {author} {\bibfnamefont {A.}~\bibnamefont {Zisserman}},\ }\href@noop {} {} (\bibinfo {year} {2014}),\ \Eprint {https://arxiv.org/abs/1312.6034} {arXiv:1312.6034} \BibitemShut {NoStop}%
\bibitem [{\citenamefont {Gazizov}\ and\ \citenamefont {Kowalski}(2005)}]{anis}%
  \BibitemOpen
  \bibfield  {author} {\bibinfo {author} {\bibfnamefont {A.}~\bibnamefont {Gazizov}}\ and\ \bibinfo {author} {\bibfnamefont {M.~P.}\ \bibnamefont {Kowalski}},\ }\href@noop {} {\bibfield  {journal} {\bibinfo  {journal} {Computer Physics Communications}\ }\textbf {\bibinfo {volume} {172}} (\bibinfo {year} {2005})}\BibitemShut {NoStop}%
\bibitem [{\citenamefont {{Honda}}\ \emph {et~al.}(2007)\citenamefont {{Honda}}, \citenamefont {{Kajita}}, \citenamefont {{Kasahara}}, \citenamefont {{Midorikawa}},\ and\ \citenamefont {{Sanuki}}}]{HKKMS06}%
  \BibitemOpen
  \bibfield  {author} {\bibinfo {author} {\bibfnamefont {M.}~\bibnamefont {{Honda}}}, \bibinfo {author} {\bibfnamefont {T.}~\bibnamefont {{Kajita}}}, \bibinfo {author} {\bibfnamefont {K.}~\bibnamefont {{Kasahara}}}, \bibinfo {author} {\bibfnamefont {S.}~\bibnamefont {{Midorikawa}}},\ and\ \bibinfo {author} {\bibfnamefont {T.}~\bibnamefont {{Sanuki}}},\ }\href@noop {} {\bibfield  {journal} {\bibinfo  {journal} {Phys. Rev.}\ }\textbf {\bibinfo {volume} {D75}},\ \bibinfo {pages} {043006} (\bibinfo {year} {2007})}\BibitemShut {NoStop}%
\bibitem [{\citenamefont {{IceCube Collaboration}}\ \emph {et~al.}(2011)\citenamefont {{IceCube Collaboration}}, \citenamefont {Abbasi} \emph {et~al.}}]{ICAtm2011}%
  \BibitemOpen
  \bibfield  {author} {\bibinfo {author} {\bibnamefont {{IceCube Collaboration}}}, \bibinfo {author} {\bibfnamefont {R.}~\bibnamefont {Abbasi}}, \emph {et~al.},\ }\href@noop {} {\bibfield  {journal} {\bibinfo  {journal} {Phys. Rev.}\ }\textbf {\bibinfo {volume} {D83}},\ \bibinfo {pages} {012001} (\bibinfo {year} {2011})}\BibitemShut {NoStop}%
\bibitem [{\citenamefont {{IceCube Collaboration}}\ \emph {et~al.}(2013{\natexlab{b}})\citenamefont {{IceCube Collaboration}}, \citenamefont {Aartsen} \emph {et~al.}}]{ICAtm2013}%
  \BibitemOpen
  \bibfield  {author} {\bibinfo {author} {\bibnamefont {{IceCube Collaboration}}}, \bibinfo {author} {\bibfnamefont {M.~G.}\ \bibnamefont {Aartsen}}, \emph {et~al.},\ }\href@noop {} {\bibfield  {journal} {\bibinfo  {journal} {Phys. Rev. Lett.}\ }\textbf {\bibinfo {volume} {110}},\ \bibinfo {pages} {151105} (\bibinfo {year} {2013}{\natexlab{b}})}\BibitemShut {NoStop}%
\bibitem [{\citenamefont {{IceCube Collaboration}}\ \emph {et~al.}(2015)\citenamefont {{IceCube Collaboration}}, \citenamefont {Aartsen} \emph {et~al.}}]{ICAtm2015}%
  \BibitemOpen
  \bibfield  {author} {\bibinfo {author} {\bibnamefont {{IceCube Collaboration}}}, \bibinfo {author} {\bibfnamefont {M.~G.}\ \bibnamefont {Aartsen}}, \emph {et~al.},\ }\href@noop {} {\bibfield  {journal} {\bibinfo  {journal} {Phys. Rev.}\ }\textbf {\bibinfo {volume} {D91}},\ \bibinfo {pages} {122004} (\bibinfo {year} {2015})}\BibitemShut {NoStop}%
\bibitem [{\citenamefont {{Bhattacharya}}\ \emph {et~al.}(2015)\citenamefont {{Bhattacharya}} \emph {et~al.}}]{BERSS}%
  \BibitemOpen
  \bibfield  {author} {\bibinfo {author} {\bibfnamefont {A.}~\bibnamefont {{Bhattacharya}}} \emph {et~al.},\ }\href@noop {} {\bibfield  {journal} {\bibinfo  {journal} {JHEP}\ }\textbf {\bibinfo {volume} {06}},\ \bibinfo {pages} {110 (2015)}}\BibitemShut {NoStop}%
\bibitem [{\citenamefont {Garzelli}\ \emph {et~al.}(2015)\citenamefont {Garzelli}, \citenamefont {Moch},\ and\ \citenamefont {Sigl}}]{Garzelli:2015psa}%
  \BibitemOpen
  \bibfield  {author} {\bibinfo {author} {\bibfnamefont {M.~V.}\ \bibnamefont {Garzelli}}, \bibinfo {author} {\bibfnamefont {S.}~\bibnamefont {Moch}},\ and\ \bibinfo {author} {\bibfnamefont {G.}~\bibnamefont {Sigl}},\ }\href {https://doi.org/10.1007/JHEP10(2015)115} {\bibfield  {journal} {\bibinfo  {journal} {JHEP}\ }\textbf {\bibinfo {volume} {10}},\ \bibinfo {pages} {115 (2015)}}\BibitemShut {NoStop}%
\bibitem [{\citenamefont {Gauld}\ \emph {et~al.}(2016)\citenamefont {Gauld}, \citenamefont {Rojo}, \citenamefont {Rottoli}, \citenamefont {Sarkar},\ and\ \citenamefont {Talbert}}]{Gauld:2015kvh}%
  \BibitemOpen
  \bibfield  {author} {\bibinfo {author} {\bibfnamefont {R.}~\bibnamefont {Gauld}}, \bibinfo {author} {\bibfnamefont {J.}~\bibnamefont {Rojo}}, \bibinfo {author} {\bibfnamefont {L.}~\bibnamefont {Rottoli}}, \bibinfo {author} {\bibfnamefont {S.}~\bibnamefont {Sarkar}},\ and\ \bibinfo {author} {\bibfnamefont {J.}~\bibnamefont {Talbert}},\ }\href {https://doi.org/10.1007/JHEP02(2016)130} {\bibfield  {journal} {\bibinfo  {journal} {JHEP}\ }\textbf {\bibinfo {volume} {02}},\ \bibinfo {pages} {130 (2016)}}\BibitemShut {NoStop}%
\bibitem [{\citenamefont {Heck}\ \emph {et~al.}(1998)\citenamefont {Heck}, \citenamefont {Knapp}, \citenamefont {Capdevielle}, \citenamefont {Schatz},\ and\ \citenamefont {Thouw}}]{Heck:1998vt}%
  \BibitemOpen
  \bibfield  {author} {\bibinfo {author} {\bibfnamefont {D.}~\bibnamefont {Heck}}, \bibinfo {author} {\bibfnamefont {J.}~\bibnamefont {Knapp}}, \bibinfo {author} {\bibfnamefont {J.~N.}\ \bibnamefont {Capdevielle}}, \bibinfo {author} {\bibfnamefont {G.}~\bibnamefont {Schatz}},\ and\ \bibinfo {author} {\bibfnamefont {T.}~\bibnamefont {Thouw}},\ }\href@noop {} {\bibfield  {journal} {\bibinfo  {journal} {FZKA-6019}\ } (\bibinfo {year} {1998})}\BibitemShut {NoStop}%
\bibitem [{\citenamefont {{van Santen}}(2014)}]{JakobThesis}%
  \BibitemOpen
  \bibfield  {author} {\bibinfo {author} {\bibfnamefont {J.}~\bibnamefont {{van Santen}}},\ }\href@noop {} {Ph.D. thesis},\ \bibinfo  {school} {University of Wisconsin-Madison} (\bibinfo {year} {2014})\BibitemShut {NoStop}%
\bibitem [{\citenamefont {Gaisser}(2012)}]{Gaisser:2011klf}%
  \BibitemOpen
  \bibfield  {author} {\bibinfo {author} {\bibfnamefont {T.~K.}\ \bibnamefont {Gaisser}},\ }\bibfield  {title} {\bibinfo {title} {{Spectrum of cosmic-ray nucleons, kaon production, and the atmospheric muon charge ratio}},\ }\href {https://doi.org/10.1016/j.astropartphys.2012.02.010} {\bibfield  {journal} {\bibinfo  {journal} {Astropart. Phys.}\ }\textbf {\bibinfo {volume} {35}},\ \bibinfo {pages} {801} (\bibinfo {year} {2012})}\BibitemShut {NoStop}%
\bibitem [{\citenamefont {{Ahn}}\ \emph {et~al.}(2009)\citenamefont {{Ahn}}, \citenamefont {{Engel}}, \citenamefont {{Gaisser}}, \citenamefont {{Lipari}},\ and\ \citenamefont {{Stanev}}}]{SIBYLL21}%
  \BibitemOpen
  \bibfield  {author} {\bibinfo {author} {\bibfnamefont {E.~J.}\ \bibnamefont {{Ahn}}}, \bibinfo {author} {\bibfnamefont {R.}~\bibnamefont {{Engel}}}, \bibinfo {author} {\bibfnamefont {T.~K.}\ \bibnamefont {{Gaisser}}}, \bibinfo {author} {\bibfnamefont {P.}~\bibnamefont {{Lipari}}},\ and\ \bibinfo {author} {\bibfnamefont {T.}~\bibnamefont {{Stanev}}},\ }\href@noop {} {\bibfield  {journal} {\bibinfo  {journal} {Phys. Rev.}\ }\textbf {\bibinfo {volume} {D80}},\ \bibinfo {pages} {094003} (\bibinfo {year} {2009})}\BibitemShut {NoStop}%
\bibitem [{\citenamefont {Schonert}\ \emph {et~al.}(2009)\citenamefont {Schonert}, \citenamefont {Gaisser}, \citenamefont {Resconi},\ and\ \citenamefont {Schulz}}]{Schonert:2008is}%
  \BibitemOpen
  \bibfield  {author} {\bibinfo {author} {\bibfnamefont {S.}~\bibnamefont {Schonert}}, \bibinfo {author} {\bibfnamefont {T.~K.}\ \bibnamefont {Gaisser}}, \bibinfo {author} {\bibfnamefont {E.}~\bibnamefont {Resconi}},\ and\ \bibinfo {author} {\bibfnamefont {O.}~\bibnamefont {Schulz}},\ }\href {https://doi.org/10.1103/PhysRevD.79.043009} {\bibfield  {journal} {\bibinfo  {journal} {Phys. Rev. D}\ }\textbf {\bibinfo {volume} {79}},\ \bibinfo {pages} {043009} (\bibinfo {year} {2009})}\BibitemShut {NoStop}%
\bibitem [{\citenamefont {Radel}\ and\ \citenamefont {Wiebusch}(2013)}]{Radel:2012ij}%
  \BibitemOpen
  \bibfield  {author} {\bibinfo {author} {\bibfnamefont {L.}~\bibnamefont {Radel}}\ and\ \bibinfo {author} {\bibfnamefont {C.}~\bibnamefont {Wiebusch}},\ }\bibfield  {title} {\bibinfo {title} {{Calculation of the Cherenkov light yield from electromagnetic cascades in ice with Geant4}},\ }\href {https://doi.org/10.1016/j.astropartphys.2013.01.015} {\bibfield  {journal} {\bibinfo  {journal} {Astropart. Phys.}\ }\textbf {\bibinfo {volume} {44}},\ \bibinfo {pages} {102} (\bibinfo {year} {2013})}\BibitemShut {NoStop}%
\bibitem [{\citenamefont {Landau}\ and\ \citenamefont {Pomeranchuk}(1953{\natexlab{a}})}]{Landau:1953gr}%
  \BibitemOpen
  \bibfield  {author} {\bibinfo {author} {\bibfnamefont {L.~D.}\ \bibnamefont {Landau}}\ and\ \bibinfo {author} {\bibfnamefont {I.}~\bibnamefont {Pomeranchuk}},\ }\bibfield  {title} {\bibinfo {title} {{Electron cascade process at very high-energies}},\ }\href@noop {} {\bibfield  {journal} {\bibinfo  {journal} {Dokl. Akad. Nauk Ser. Fiz.}\ }\textbf {\bibinfo {volume} {92}},\ \bibinfo {pages} {735} (\bibinfo {year} {1953}{\natexlab{a}})}\BibitemShut {NoStop}%
\bibitem [{\citenamefont {Landau}\ and\ \citenamefont {Pomeranchuk}(1953{\natexlab{b}})}]{Landau:1953um}%
  \BibitemOpen
  \bibfield  {author} {\bibinfo {author} {\bibfnamefont {L.~D.}\ \bibnamefont {Landau}}\ and\ \bibinfo {author} {\bibfnamefont {I.}~\bibnamefont {Pomeranchuk}},\ }\bibfield  {title} {\bibinfo {title} {{Limits of applicability of the theory of bremsstrahlung electrons and pair production at high-energies}},\ }\href@noop {} {\bibfield  {journal} {\bibinfo  {journal} {Dokl. Akad. Nauk Ser. Fiz.}\ }\textbf {\bibinfo {volume} {92}},\ \bibinfo {pages} {535} (\bibinfo {year} {1953}{\natexlab{b}})}\BibitemShut {NoStop}%
\bibitem [{\citenamefont {Migdal}(1956)}]{Migdal:1956tc}%
  \BibitemOpen
  \bibfield  {author} {\bibinfo {author} {\bibfnamefont {A.~B.}\ \bibnamefont {Migdal}},\ }\bibfield  {title} {\bibinfo {title} {{Bremsstrahlung and pair production in condensed media at high-energies}},\ }\href {https://doi.org/10.1103/PhysRev.103.1811} {\bibfield  {journal} {\bibinfo  {journal} {Phys. Rev.}\ }\textbf {\bibinfo {volume} {103}},\ \bibinfo {pages} {1811} (\bibinfo {year} {1956})}\BibitemShut {NoStop}%
\bibitem [{\citenamefont {Cooper-Sarkar}\ \emph {et~al.}(2011)\citenamefont {Cooper-Sarkar}, \citenamefont {Mertsch},\ and\ \citenamefont {Sarkar}}]{sarkar}%
  \BibitemOpen
  \bibfield  {author} {\bibinfo {author} {\bibfnamefont {A.}~\bibnamefont {Cooper-Sarkar}}, \bibinfo {author} {\bibfnamefont {P.}~\bibnamefont {Mertsch}},\ and\ \bibinfo {author} {\bibfnamefont {S.}~\bibnamefont {Sarkar}},\ }\href@noop {} {\bibfield  {journal} {\bibinfo  {journal} {JHEP}\ }\textbf {\bibinfo {volume} {08}},\ \bibinfo {pages} {042 (2011)}}\BibitemShut {NoStop}%
\bibitem [{\citenamefont {Aad}\ \emph {et~al.}(2022)\citenamefont {Aad} \emph {et~al.}}]{ATLAS:2021vod}%
  \BibitemOpen
  \bibfield  {author} {\bibinfo {author} {\bibfnamefont {G.}~\bibnamefont {Aad}} \emph {et~al.} (\bibinfo {collaboration} {ATLAS}),\ }\bibfield  {title} {\bibinfo {title} {{Determination of the parton distribution functions of the proton using diverse ATLAS data from $pp$ collisions at $\sqrt{s} = 7$, 8 and 13~TeV}},\ }\href {https://doi.org/10.1140/epjc/s10052-022-10217-z} {\bibfield  {journal} {\bibinfo  {journal} {Eur. Phys. J. C}\ }\textbf {\bibinfo {volume} {82}},\ \bibinfo {pages} {438} (\bibinfo {year} {2022})}\BibitemShut {NoStop}%
\bibitem [{\citenamefont {Radescu}(2010)}]{Radescu:2010zz}%
  \BibitemOpen
  \bibfield  {author} {\bibinfo {author} {\bibfnamefont {V.}~\bibnamefont {Radescu}} (\bibinfo {collaboration} {H1, ZEUS}),\ }\href {https://doi.org/10.22323/1.120.0168} {\bibfield  {journal} {\bibinfo  {journal} {PoS ICHEP2010}\ ,\ \bibinfo {pages} {168}} (\bibinfo {year} {2010})}\BibitemShut {NoStop}%
\bibitem [{\citenamefont {Hou}\ \emph {et~al.}(2021)\citenamefont {Hou} \emph {et~al.}}]{Hou:2019efy}%
  \BibitemOpen
  \bibfield  {author} {\bibinfo {author} {\bibfnamefont {T.-J.}\ \bibnamefont {Hou}} \emph {et~al.},\ }\href {https://doi.org/10.1103/PhysRevD.103.014013} {\bibfield  {journal} {\bibinfo  {journal} {Phys. Rev. D}\ }\textbf {\bibinfo {volume} {103}},\ \bibinfo {pages} {014013} (\bibinfo {year} {2021})}\BibitemShut {NoStop}%
\bibitem [{\citenamefont {Bailey}\ \emph {et~al.}(2021)\citenamefont {Bailey}, \citenamefont {Cridge}, \citenamefont {Harland-Lang}, \citenamefont {Martin},\ and\ \citenamefont {Thorne}}]{Bailey:2020ooq}%
  \BibitemOpen
  \bibfield  {author} {\bibinfo {author} {\bibfnamefont {S.}~\bibnamefont {Bailey}}, \bibinfo {author} {\bibfnamefont {T.}~\bibnamefont {Cridge}}, \bibinfo {author} {\bibfnamefont {L.~A.}\ \bibnamefont {Harland-Lang}}, \bibinfo {author} {\bibfnamefont {A.~D.}\ \bibnamefont {Martin}},\ and\ \bibinfo {author} {\bibfnamefont {R.~S.}\ \bibnamefont {Thorne}},\ }\href {https://doi.org/10.1140/epjc/s10052-021-09057-0} {\bibfield  {journal} {\bibinfo  {journal} {Eur. Phys. J. C}\ }\textbf {\bibinfo {volume} {81}},\ \bibinfo {pages} {341} (\bibinfo {year} {2021})}\BibitemShut {NoStop}%
\bibitem [{\citenamefont {Faura}\ \emph {et~al.}(2020)\citenamefont {Faura}, \citenamefont {Iranipour}, \citenamefont {Nocera}, \citenamefont {Rojo},\ and\ \citenamefont {Ubiali}}]{Faura:2020oom}%
  \BibitemOpen
  \bibfield  {author} {\bibinfo {author} {\bibfnamefont {F.}~\bibnamefont {Faura}}, \bibinfo {author} {\bibfnamefont {S.}~\bibnamefont {Iranipour}}, \bibinfo {author} {\bibfnamefont {E.~R.}\ \bibnamefont {Nocera}}, \bibinfo {author} {\bibfnamefont {J.}~\bibnamefont {Rojo}},\ and\ \bibinfo {author} {\bibfnamefont {M.}~\bibnamefont {Ubiali}},\ }\href {https://doi.org/10.1140/epjc/s10052-020-08749-3} {\bibfield  {journal} {\bibinfo  {journal} {Eur. Phys. J. C}\ }\textbf {\bibinfo {volume} {80}},\ \bibinfo {pages} {1168} (\bibinfo {year} {2020})}\BibitemShut {NoStop}%
\bibitem [{\citenamefont {Aaboud}\ \emph {et~al.}(2017)\citenamefont {Aaboud} \emph {et~al.}}]{ATLAS:2016nqi}%
  \BibitemOpen
  \bibfield  {author} {\bibinfo {author} {\bibfnamefont {M.}~\bibnamefont {Aaboud}} \emph {et~al.} (\bibinfo {collaboration} {ATLAS}),\ }\href {https://doi.org/10.1140/epjc/s10052-017-4911-9} {\bibfield  {journal} {\bibinfo  {journal} {Eur. Phys. J. C}\ }\textbf {\bibinfo {volume} {77}},\ \bibinfo {pages} {367} (\bibinfo {year} {2017})}\BibitemShut {NoStop}%
\bibitem [{\citenamefont {Zhou}\ and\ \citenamefont {Beacom}(2020)}]{Zhou:2019frk}%
  \BibitemOpen
  \bibfield  {author} {\bibinfo {author} {\bibfnamefont {B.}~\bibnamefont {Zhou}}\ and\ \bibinfo {author} {\bibfnamefont {J.~F.}\ \bibnamefont {Beacom}},\ }\href {https://doi.org/10.1103/PhysRevD.101.036010} {\bibfield  {journal} {\bibinfo  {journal} {Phys. Rev. D}\ }\textbf {\bibinfo {volume} {101}},\ \bibinfo {pages} {036010} (\bibinfo {year} {2020})}\BibitemShut {NoStop}%
\bibitem [{\citenamefont {Soto}\ \emph {et~al.}(2022)\citenamefont {Soto}, \citenamefont {Zhelnin}, \citenamefont {Safa},\ and\ \citenamefont {Arg\"uelles}}]{Soto:2021vdc}%
  \BibitemOpen
  \bibfield  {author} {\bibinfo {author} {\bibfnamefont {A.~G.}\ \bibnamefont {Soto}}, \bibinfo {author} {\bibfnamefont {P.}~\bibnamefont {Zhelnin}}, \bibinfo {author} {\bibfnamefont {I.}~\bibnamefont {Safa}},\ and\ \bibinfo {author} {\bibfnamefont {C.~A.}\ \bibnamefont {Arg\"uelles}},\ }\href {https://doi.org/10.1103/PhysRevLett.128.171101} {\bibfield  {journal} {\bibinfo  {journal} {Phys. Rev. Lett.}\ }\textbf {\bibinfo {volume} {128}},\ \bibinfo {pages} {171101} (\bibinfo {year} {2022})}\BibitemShut {NoStop}%
\bibitem [{\citenamefont {Feldman}\ and\ \citenamefont {Cousins}(1998)}]{Feldman:1997qc}%
  \BibitemOpen
  \bibfield  {author} {\bibinfo {author} {\bibfnamefont {G.~J.}\ \bibnamefont {Feldman}}\ and\ \bibinfo {author} {\bibfnamefont {R.~D.}\ \bibnamefont {Cousins}},\ }\href {https://doi.org/10.1103/PhysRevD.57.3873} {\bibfield  {journal} {\bibinfo  {journal} {Phys. Rev. D}\ }\textbf {\bibinfo {volume} {57}},\ \bibinfo {pages} {3873} (\bibinfo {year} {1998})}\BibitemShut {NoStop}%
\bibitem [{\citenamefont {Lu}(2018)}]{Lu:2017nti}%
  \BibitemOpen
  \bibfield  {author} {\bibinfo {author} {\bibfnamefont {L.}~\bibnamefont {Lu}} (\bibinfo {collaboration} {IceCube}),\ }\href {https://doi.org/10.22323/1.301.1002} {\bibfield  {journal} {\bibinfo  {journal} {PoS ICRC2017}\ ,\ \bibinfo {pages} {1002}} (\bibinfo {year} {2018})}\BibitemShut {NoStop}%
\bibitem [{\citenamefont {Huennefeld}\ \emph {et~al.}(2021)\citenamefont {Huennefeld} \emph {et~al.}}]{IceCube:2021umt}%
  \BibitemOpen
  \bibfield  {author} {\bibinfo {author} {\bibfnamefont {M.}~\bibnamefont {Huennefeld}} \emph {et~al.} (\bibinfo {collaboration} {IceCube}),\ }\href {https://doi.org/10.22323/1.395.1065} {\bibfield  {journal} {\bibinfo  {journal} {PoS ICRC2021}\ ,\ \bibinfo {pages} {1065}} (\bibinfo {year} {2021})}\BibitemShut {NoStop}%
\bibitem [{\citenamefont {Halzen}\ and\ \citenamefont {Saltzberg}(1998)}]{Halzen:1998be}%
  \BibitemOpen
  \bibfield  {author} {\bibinfo {author} {\bibfnamefont {F.}~\bibnamefont {Halzen}}\ and\ \bibinfo {author} {\bibfnamefont {D.}~\bibnamefont {Saltzberg}},\ }\href {https://doi.org/10.1103/PhysRevLett.81.4305} {\bibfield  {journal} {\bibinfo  {journal} {Phys. Rev. Lett.}\ }\textbf {\bibinfo {volume} {81}},\ \bibinfo {pages} {4305} (\bibinfo {year} {1998})}\BibitemShut {NoStop}%
\bibitem [{\citenamefont {Moosavi-Dezfooli}\ \emph {et~al.}(2016)\citenamefont {Moosavi-Dezfooli}, \citenamefont {Fawzi},\ and\ \citenamefont {Frossard}}]{moosavi2016deepfool}%
  \BibitemOpen
  \bibfield  {author} {\bibinfo {author} {\bibfnamefont {S.-M.}\ \bibnamefont {Moosavi-Dezfooli}}, \bibinfo {author} {\bibfnamefont {A.}~\bibnamefont {Fawzi}},\ and\ \bibinfo {author} {\bibfnamefont {P.}~\bibnamefont {Frossard}},\ }\bibfield  {title} {\bibinfo {title} {Deepfool: a simple and accurate method to fool deep neural networks},\ }in\ \href@noop {} {\emph {\bibinfo {booktitle} {Proceedings of the IEEE conference on computer vision and pattern recognition}}}\ (\bibinfo {year} {2016})\ pp.\ \bibinfo {pages} {2574--2582}\BibitemShut {NoStop}%
\bibitem [{\citenamefont {{IceCube Collaboration}}\ \emph {et~al.}(2013{\natexlab{c}})\citenamefont {{IceCube Collaboration}}, \citenamefont {Aartsen} \emph {et~al.}}]{ehe-prl-2013}%
  \BibitemOpen
  \bibfield  {author} {\bibinfo {author} {\bibnamefont {{IceCube Collaboration}}}, \bibinfo {author} {\bibfnamefont {M.~G.}\ \bibnamefont {Aartsen}}, \emph {et~al.},\ }\href@noop {} {\bibfield  {journal} {\bibinfo  {journal} {Phys. Rev. Lett.}\ }\textbf {\bibinfo {volume} {111}},\ \bibinfo {pages} {021103} (\bibinfo {year} {2013}{\natexlab{c}})}\BibitemShut {NoStop}%
\bibitem [{\citenamefont {Aartsen}\ \emph {et~al.}(2013{\natexlab{b}})\citenamefont {Aartsen} \emph {et~al.}}]{IceCube:2013llx}%
  \BibitemOpen
  \bibfield  {author} {\bibinfo {author} {\bibfnamefont {M.~G.}\ \bibnamefont {Aartsen}} \emph {et~al.} (\bibinfo {collaboration} {IceCube}),\ }\bibfield  {title} {\bibinfo {title} {{Measurement of South Pole ice transparency with the IceCube LED calibration system}},\ }\href {https://doi.org/10.1016/j.nima.2013.01.054} {\bibfield  {journal} {\bibinfo  {journal} {Nucl. Instrum. Meth. A}\ }\textbf {\bibinfo {volume} {711}},\ \bibinfo {pages} {73} (\bibinfo {year} {2013}{\natexlab{b}})}\BibitemShut {NoStop}%
\bibitem [{\citenamefont {Chirkin}(2013)}]{Chirkin:2013lpu}%
  \BibitemOpen
  \bibfield  {author} {\bibinfo {author} {\bibfnamefont {D.}~\bibnamefont {Chirkin}} (\bibinfo {collaboration} {IceCube}),\ }in\ \href@noop {} {\emph {\bibinfo {booktitle} {{33rd International Cosmic Ray Conference}}}}\ (\bibinfo {year} {2013})\ p.\ \bibinfo {pages} {0580}\BibitemShut {NoStop}%
\bibitem [{\citenamefont {{Kuiper}}(1960)}]{Kuiper}%
  \BibitemOpen
  \bibfield  {author} {\bibinfo {author} {\bibfnamefont {N.}~\bibnamefont {{Kuiper}}},\ }\href@noop {} {\bibfield  {journal} {\bibinfo  {journal} {Nederl Akad Wetensch Proc Ser A.}\ ,\ \bibinfo {pages} {38}} (\bibinfo {year} {1960})}\BibitemShut {NoStop}%
\bibitem [{\citenamefont {Ranft}(1995)}]{DPMJET255}%
  \BibitemOpen
  \bibfield  {author} {\bibinfo {author} {\bibfnamefont {J.}~\bibnamefont {Ranft}},\ }\href {https://doi.org/10.1103/PhysRevD.51.64} {\bibfield  {journal} {\bibinfo  {journal} {Phys. Rev. D}\ }\textbf {\bibinfo {volume} {51}},\ \bibinfo {pages} {64} (\bibinfo {year} {1995})}\BibitemShut {NoStop}%
\bibitem [{\citenamefont {{Kolmogorov}}(1933)}]{KS}%
  \BibitemOpen
  \bibfield  {author} {\bibinfo {author} {\bibfnamefont {A.}~\bibnamefont {{Kolmogorov}}},\ }\href@noop {} {\bibfield  {journal} {\bibinfo  {journal} {G. Ist. Ital. Attuari.}\ }\textbf {\bibinfo {volume} {4}},\ \bibinfo {pages} {83} (\bibinfo {year} {1933})}\BibitemShut {NoStop}%
\bibitem [{\citenamefont {Abbasi}\ \emph {et~al.}(2021{\natexlab{b}})\citenamefont {Abbasi} \emph {et~al.}}]{IceCube:2020tcq}%
  \BibitemOpen
  \bibfield  {author} {\bibinfo {author} {\bibfnamefont {R.}~\bibnamefont {Abbasi}} \emph {et~al.} (\bibinfo {collaboration} {IceCube}),\ }\bibfield  {title} {\bibinfo {title} {{LeptonInjector and LeptonWeighter: A neutrino event generator and weighter for neutrino observatories}},\ }\href {https://doi.org/10.1016/j.cpc.2021.108018} {\bibfield  {journal} {\bibinfo  {journal} {Comput. Phys. Commun.}\ }\textbf {\bibinfo {volume} {266}},\ \bibinfo {pages} {108018} (\bibinfo {year} {2021}{\natexlab{b}})}\BibitemShut {NoStop}%
\bibitem [{\citenamefont {Aartsen}\ \emph {et~al.}(2016{\natexlab{b}})\citenamefont {Aartsen} \emph {et~al.}}]{IceCube:2015wro}%
  \BibitemOpen
  \bibfield  {author} {\bibinfo {author} {\bibfnamefont {M.~G.}\ \bibnamefont {Aartsen}} \emph {et~al.} (\bibinfo {collaboration} {IceCube}),\ }\href {https://doi.org/10.1016/j.astropartphys.2016.01.006} {\bibfield  {journal} {\bibinfo  {journal} {Astropart. Phys.}\ }\textbf {\bibinfo {volume} {78}},\ \bibinfo {pages} {1} (\bibinfo {year} {2016}{\natexlab{b}})}\BibitemShut {NoStop}%
\bibitem [{\citenamefont {Agafonova}\ \emph {et~al.}(2019{\natexlab{b}})\citenamefont {Agafonova} \emph {et~al.}}]{LVD:2019zlh}%
  \BibitemOpen
  \bibfield  {author} {\bibinfo {author} {\bibfnamefont {N.~Y.}\ \bibnamefont {Agafonova}} \emph {et~al.} (\bibinfo {collaboration} {LVD}),\ }\href {https://doi.org/10.1103/PhysRevD.100.062002} {\bibfield  {journal} {\bibinfo  {journal} {Phys. Rev. D}\ }\textbf {\bibinfo {volume} {100}},\ \bibinfo {pages} {062002} (\bibinfo {year} {2019}{\natexlab{b}})}\BibitemShut {NoStop}%
\bibitem [{\citenamefont {Bogdanov}\ \emph {et~al.}(2012)\citenamefont {Bogdanov}, \citenamefont {Kokoulin}, \citenamefont {Novoseltsev}, \citenamefont {Novoseltseva}, \citenamefont {Petkov},\ and\ \citenamefont {Petrukhin}}]{Bogdanov:2009ny}%
  \BibitemOpen
  \bibfield  {author} {\bibinfo {author} {\bibfnamefont {A.~G.}\ \bibnamefont {Bogdanov}}, \bibinfo {author} {\bibfnamefont {R.~P.}\ \bibnamefont {Kokoulin}}, \bibinfo {author} {\bibfnamefont {Y.~F.}\ \bibnamefont {Novoseltsev}}, \bibinfo {author} {\bibfnamefont {R.~V.}\ \bibnamefont {Novoseltseva}}, \bibinfo {author} {\bibfnamefont {V.~B.}\ \bibnamefont {Petkov}},\ and\ \bibinfo {author} {\bibfnamefont {A.~A.}\ \bibnamefont {Petrukhin}},\ }\href {https://doi.org/10.1016/j.astropartphys.2012.06.004} {\bibfield  {journal} {\bibinfo  {journal} {Astropart. Phys.}\ }\textbf {\bibinfo {volume} {36}},\ \bibinfo {pages} {224} (\bibinfo {year} {2012})}\BibitemShut {NoStop}%
\bibitem [{\citenamefont {Chirkin}\ and\ \citenamefont {Rongen}(2019)}]{chirkin2019light}%
  \BibitemOpen
  \bibfield  {author} {\bibinfo {author} {\bibfnamefont {D.}~\bibnamefont {Chirkin}}\ and\ \bibinfo {author} {\bibfnamefont {M.}~\bibnamefont {Rongen}},\ }\href@noop {} {\bibinfo {title} {Light diffusion in birefringent polycrystals and the icecube ice anisotropy}} (\bibinfo {year} {2019}),\ \Eprint {https://arxiv.org/abs/1908.07608} {arXiv:1908.07608 [astro-ph.HE]} \BibitemShut {NoStop}%
\end{thebibliography}

%

\appendix
\newpage
\onecolumngrid
\renewcommand{\thetable}{A\arabic{table}}
\setcounter{table}{0}  
\renewcommand{\thefigure}{A\arabic{figure}}
\setcounter{figure}{0}
\renewcommand{\theequation}{A\arabic{equation}}
\setcounter{equation}{0}
~\\

\begin{large}\begin{center}
                \textbf{Appendix}
\end{center}\end{large}

The following appendix includes event displays and saliency maps not shown in the main text, and a more detailed discussion of backgrounds and data-driven studies of the CNN performance.

\newpage
\section{Event Displays}
\begin{figure}[ht]
\centering
\begin{tabular}{cc}
    \includegraphics[width=0.35\linewidth]{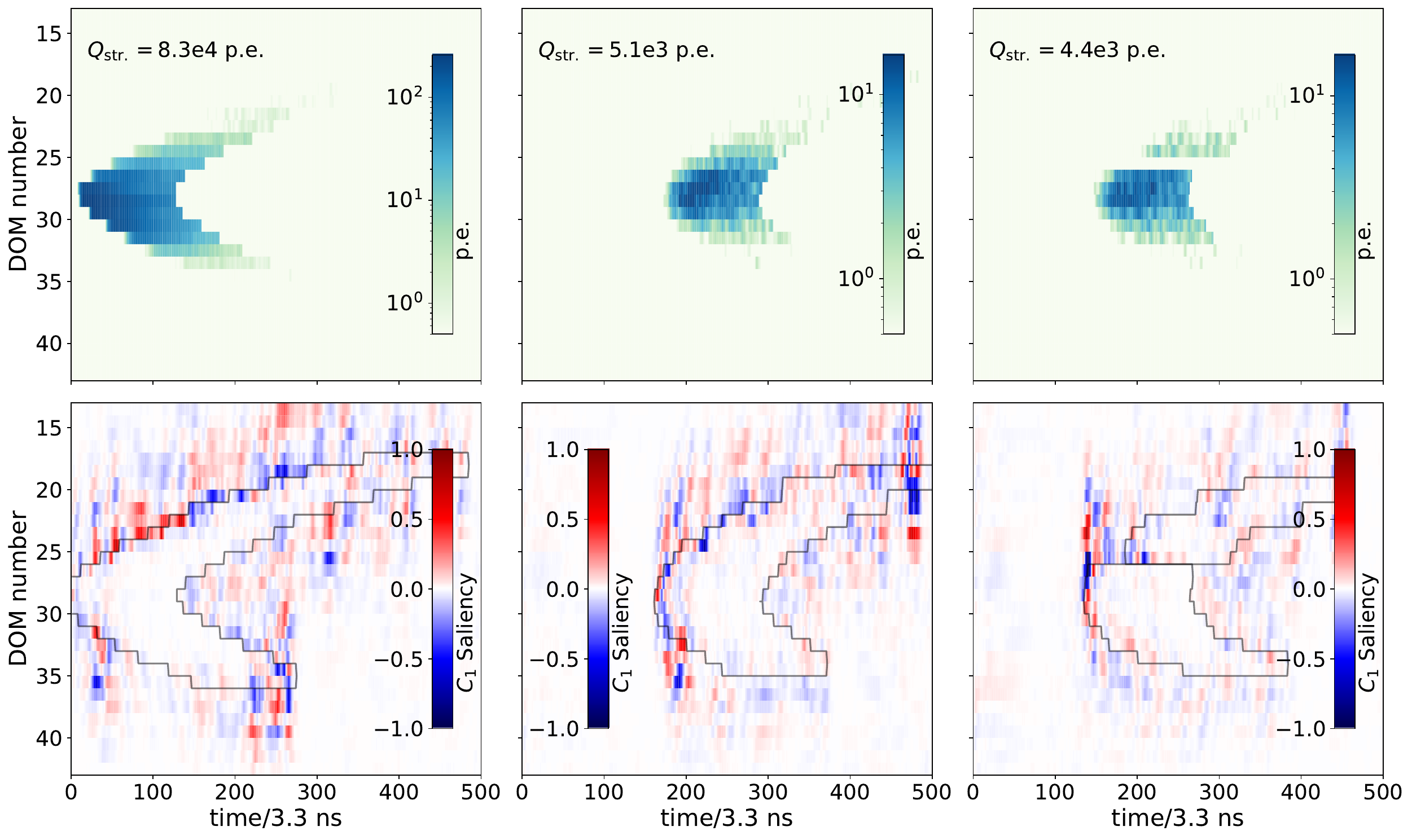}&
    \includegraphics[width=0.35\linewidth]{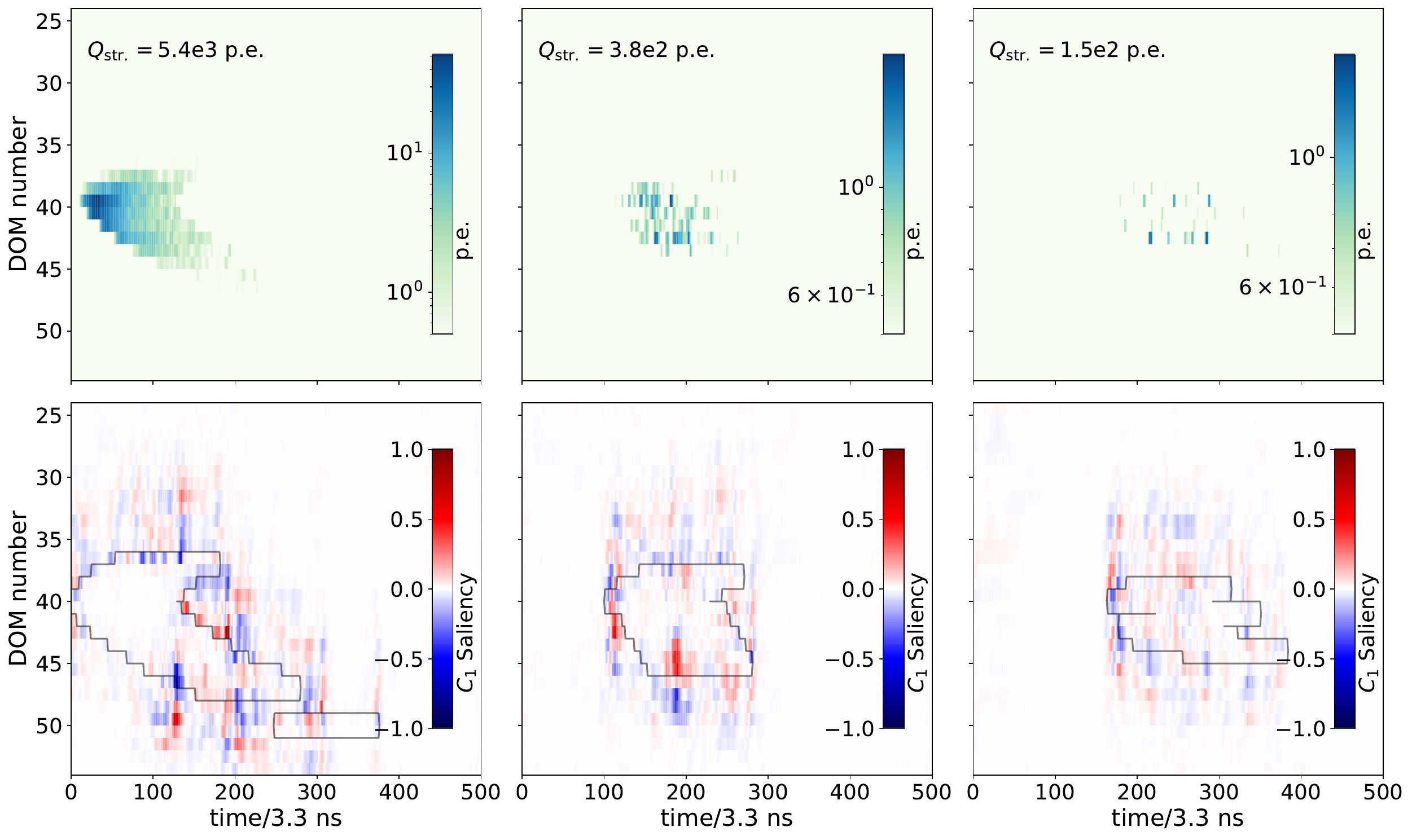}\\
    \includegraphics[width=0.35\linewidth]{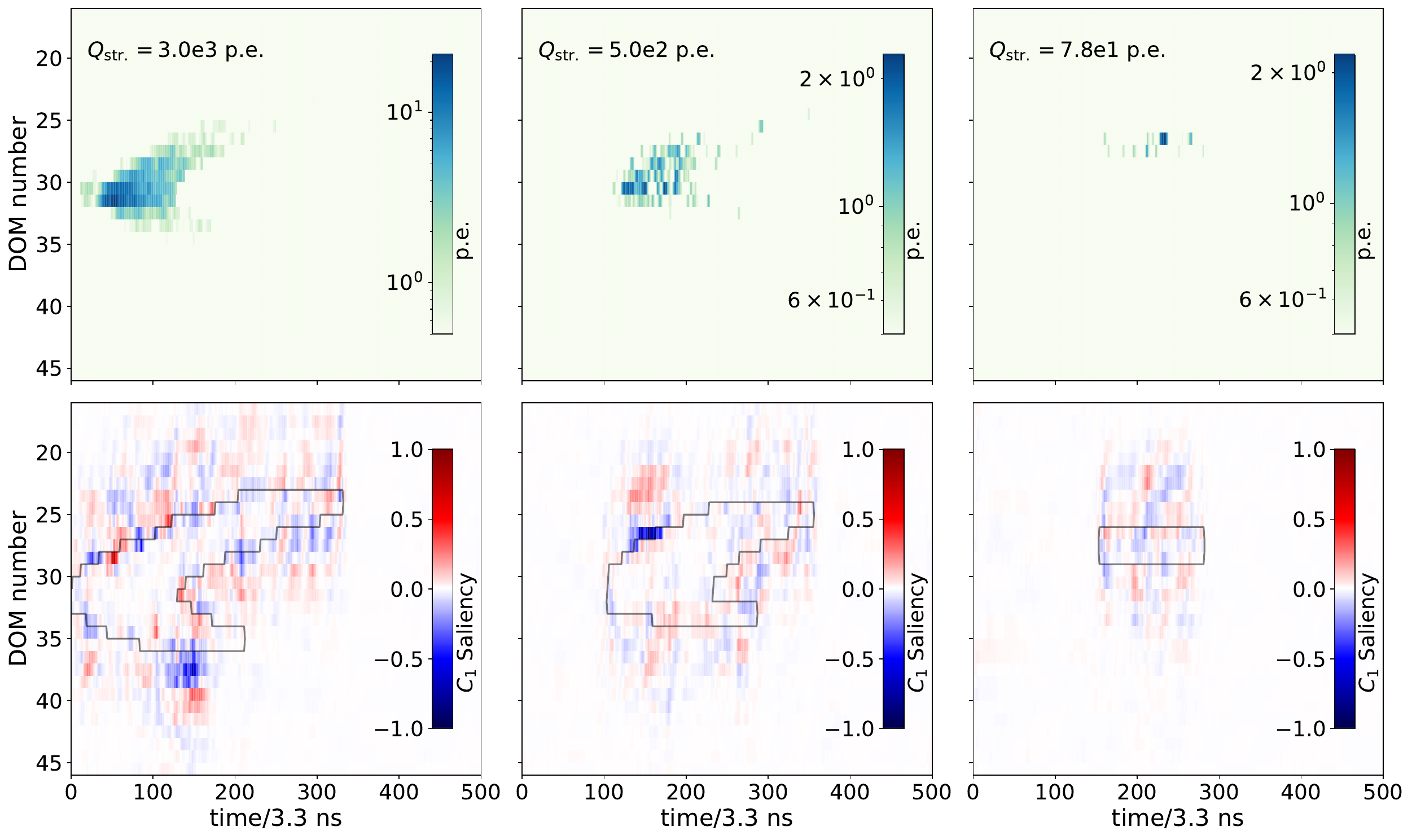}&
    \includegraphics[width=0.35\linewidth]{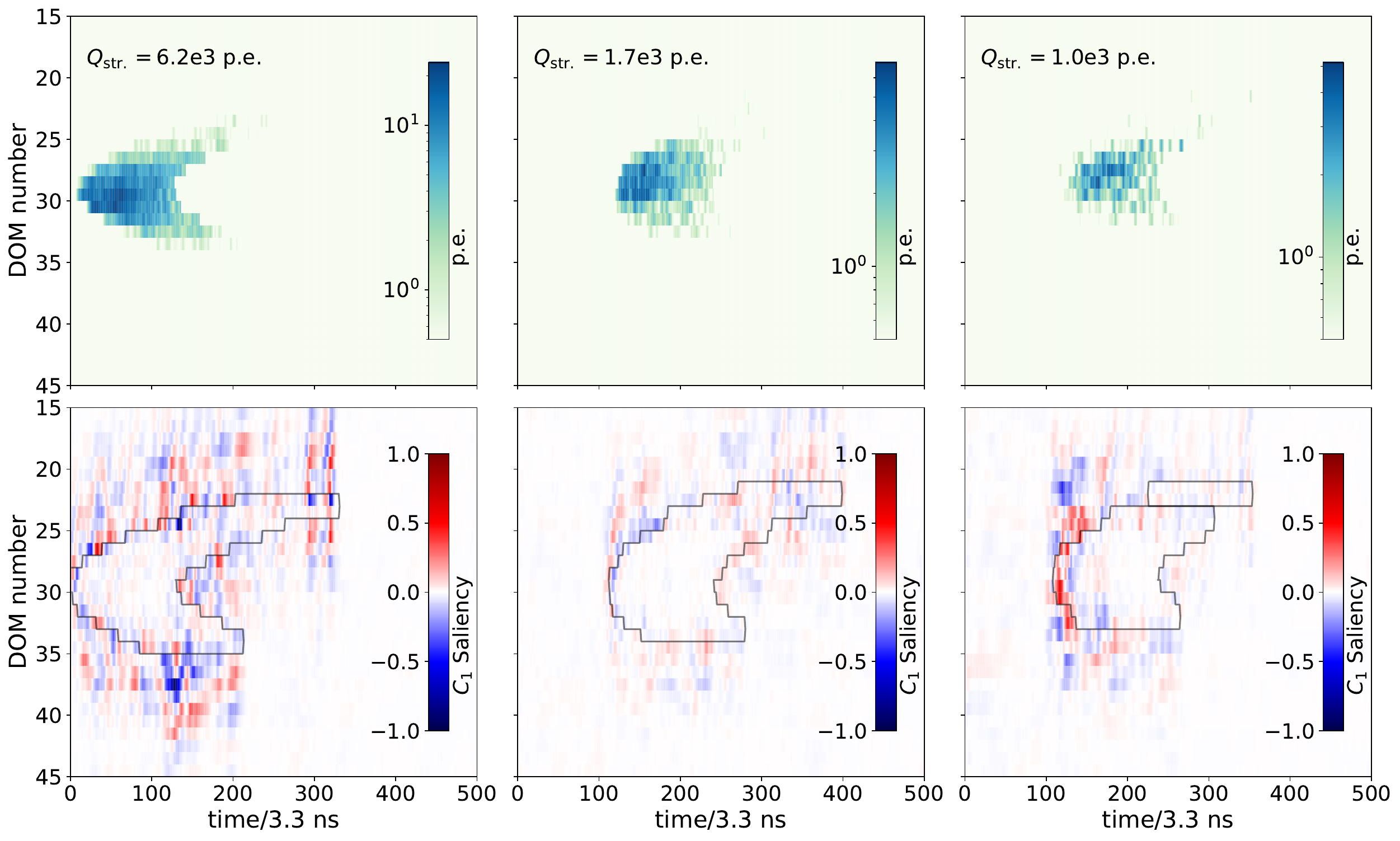}\\
    \includegraphics[width=0.35\linewidth]{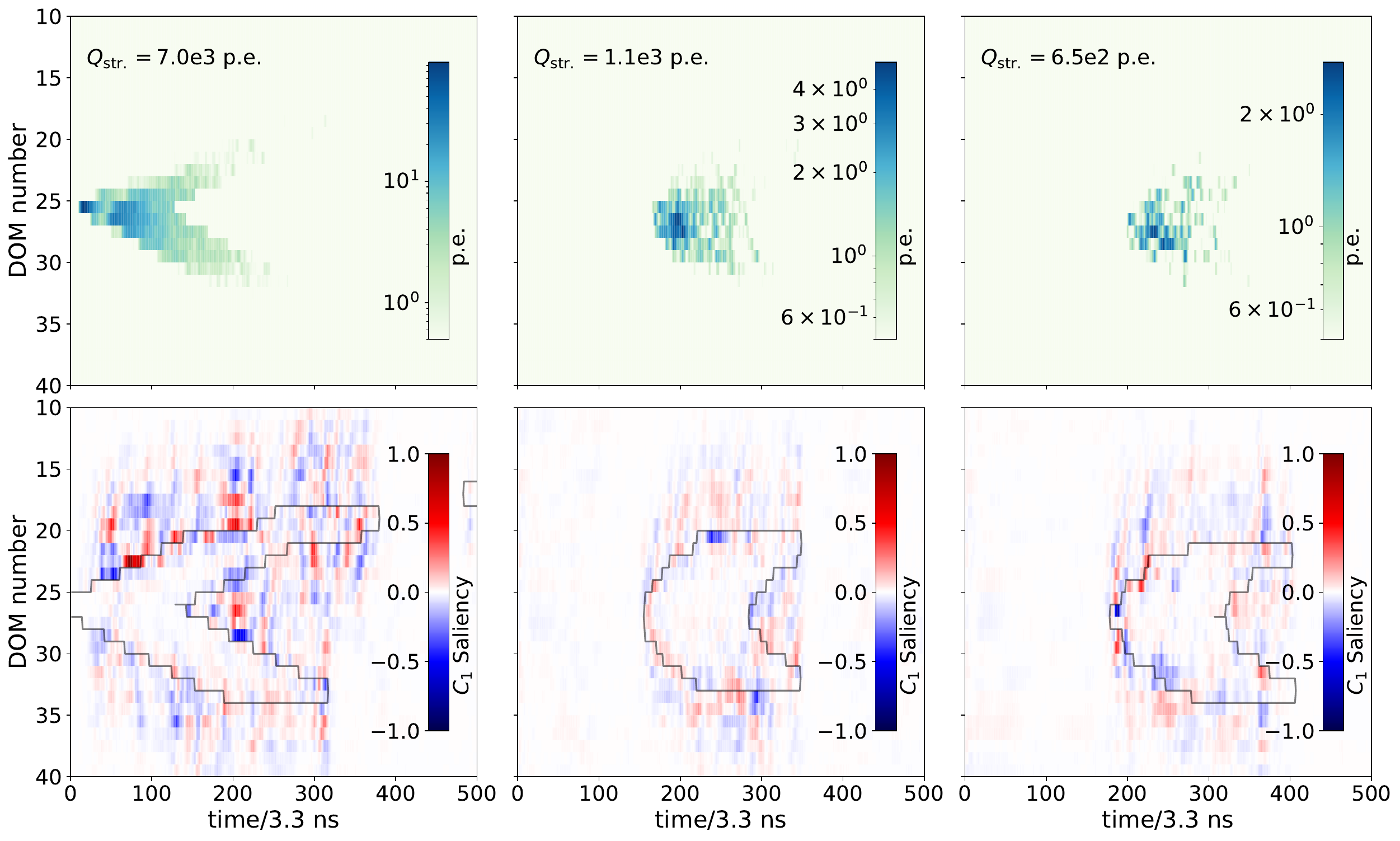}&
    \includegraphics[width=0.35\linewidth]{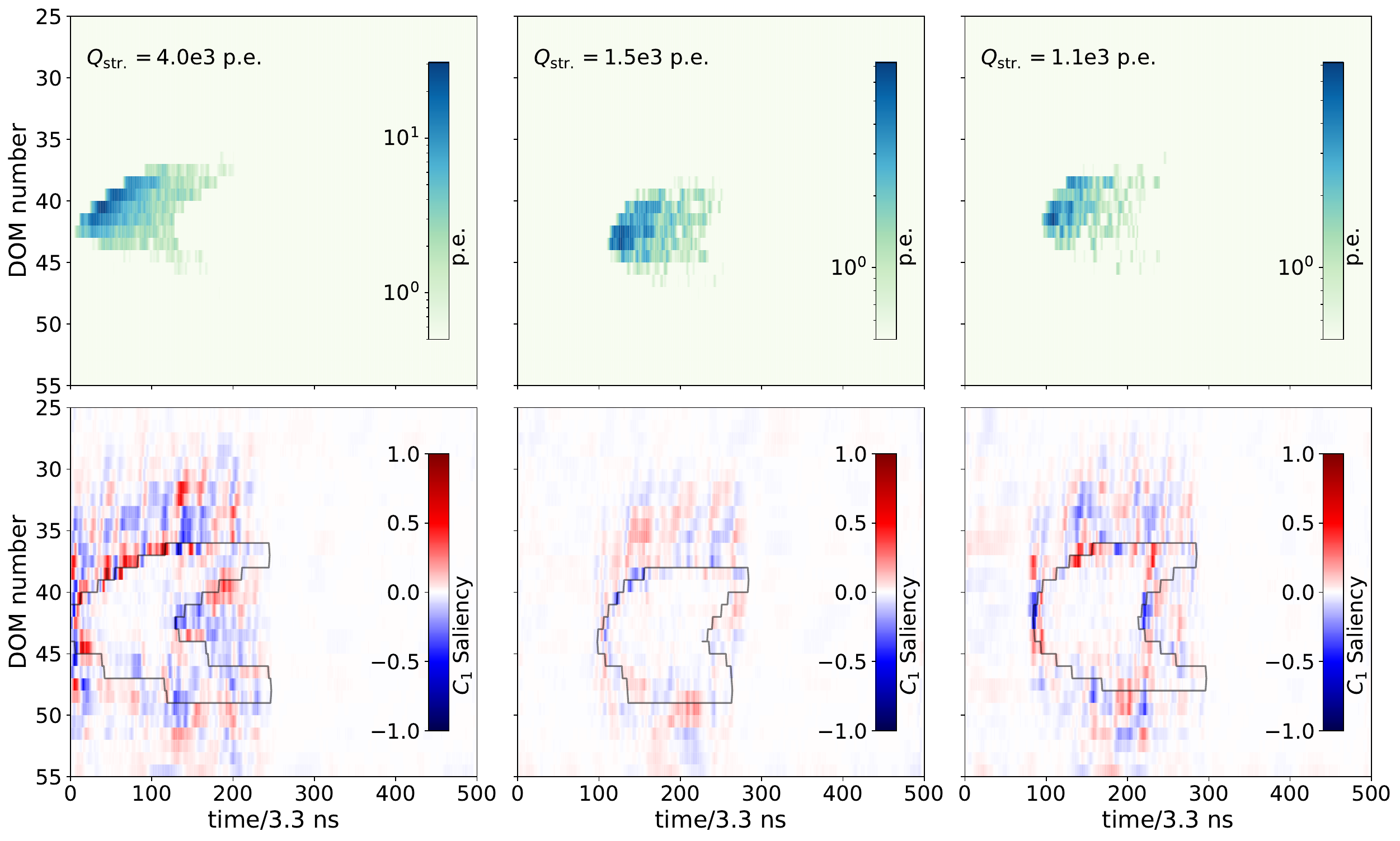}\\
    \includegraphics[width=0.35\linewidth]{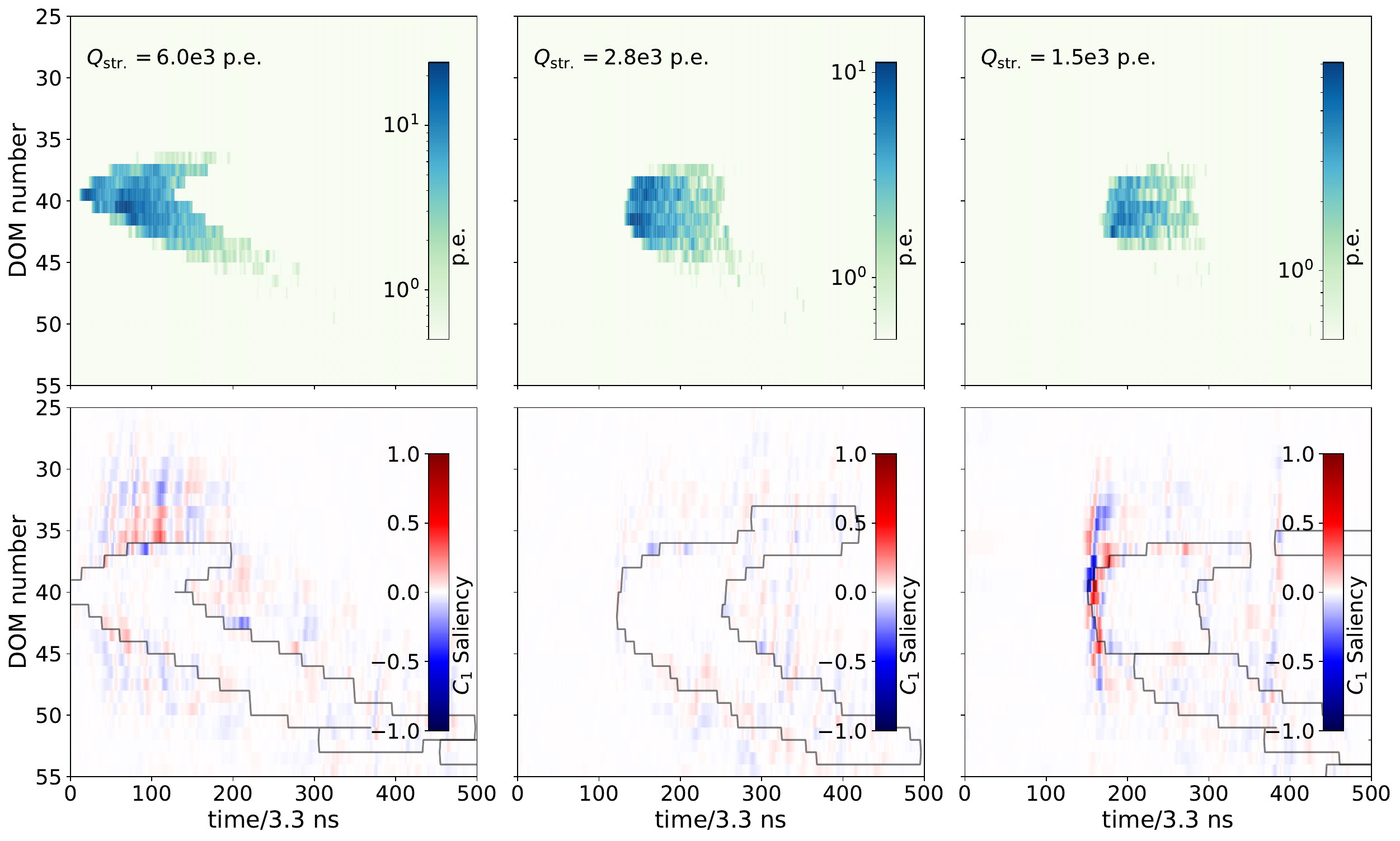}
\end{tabular}
\caption{The figures show the 2-d images and saliency maps for all candidate $\nu_\tau^\mathrm{astro}$ events.  The three columns in each figure correspond to the three strings in the selected event.  The top row in each figure shows the measured light level as a function of DOM number (proportional to depth) and time (in 3.3~ns bins).  These $(60 \times 500)$-pixel images from simulated signal and background were used to train the CNNs.  The bottom row in each figure shows the saliency, scaled from [-1,1], with red (blue) regions indicating where increased (decreased) light would increase $C_1$ (see text).  The contour (solid line) superimposed on the saliency plots corresponds to the pixels where the light level went to zero, and is roughly an outline of the light-level plot above it. 
The events depicted were detected in Jan. 2012 (top left), Jul. 2013 (top right), Oct. 2013 (second row left), Dec. 2014 (second row right), Apr. 2015 (third row left), Sep. 2015 (third row right) and Nov. 2019 (bottom left).  (In the top left figure, one of the DOMs in the third string is faulty and had been removed from the data stream, resulting in the blank horizontal region visible in the figure.  As this is a very rare occurrence, the CNN was not trained with data that included it, but its absence did not have a noticeable impact on the CNN scores.)}
\label{fig:SevenEventDisplays}
\end{figure}

Images and saliency maps for all seven candidate $\nu_\tau^\mathrm{astro}$ events are shown in Fig.~\ref{fig:SevenEventDisplays}.

\clearpage

\section{Candidate Event Spatial Distribution}

\begin{figure}[h]
\centering

\begin{minipage}{0.45\textwidth}
\centering
    \includegraphics[width=1.0\textwidth]{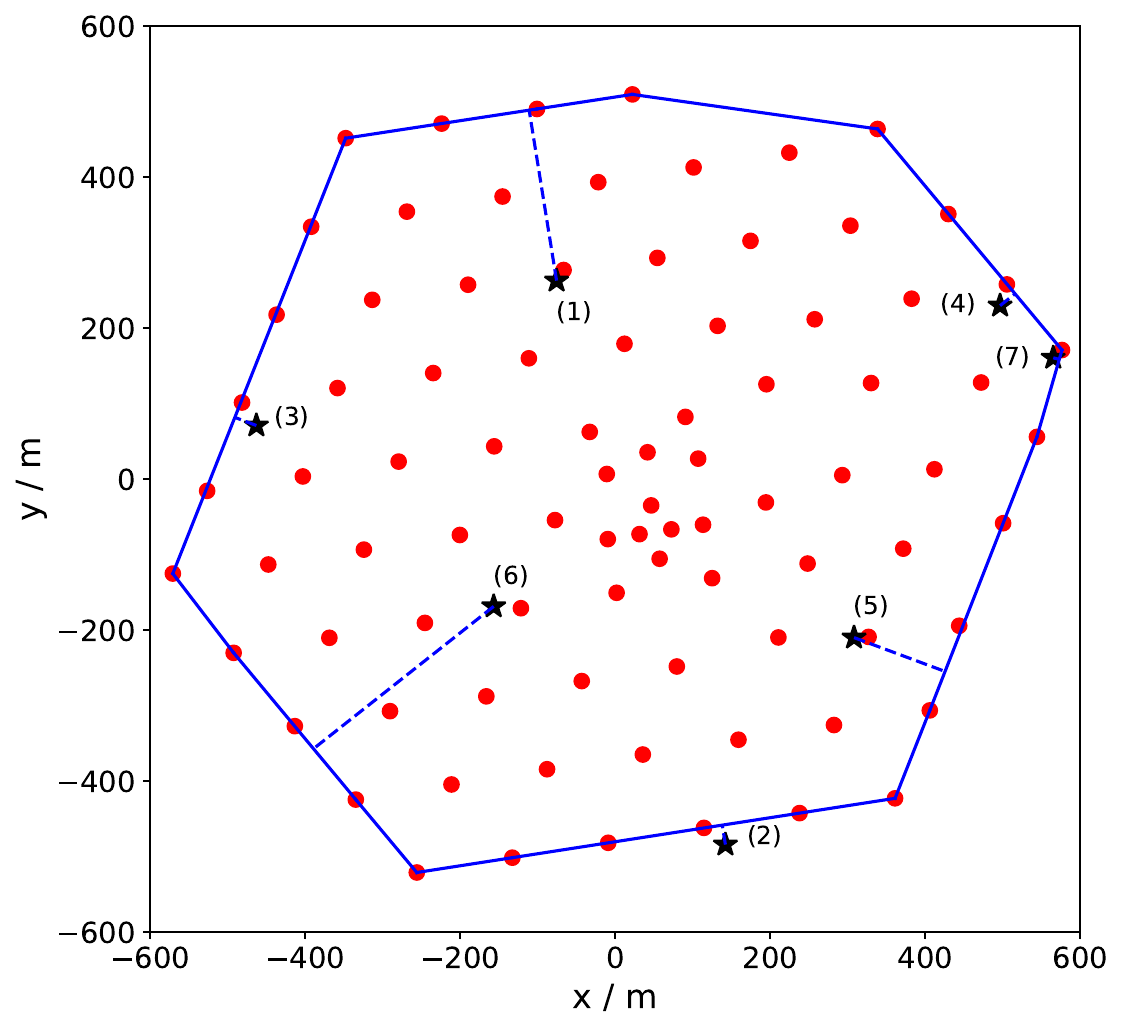}
    \caption{Top view of the reconstructed vertex positions~\protect\cite{IceCube:2021umt} of the seven $\nu_\tau^\mathrm{astro}$ candidate events, indicated by stars, with the positions of IceCube strings shown as circles.  The dashed lines show the horizontal distance from the edge of the detector (delineated by the solid line) to the event vertex.  (The numbers near each data point correspond to the events; see text.)}
    \label{fig:HullDistances}
\end{minipage}
\hfill
\begin{minipage}{0.45\textwidth}
\centering
    \includegraphics[width=1.0\textwidth]{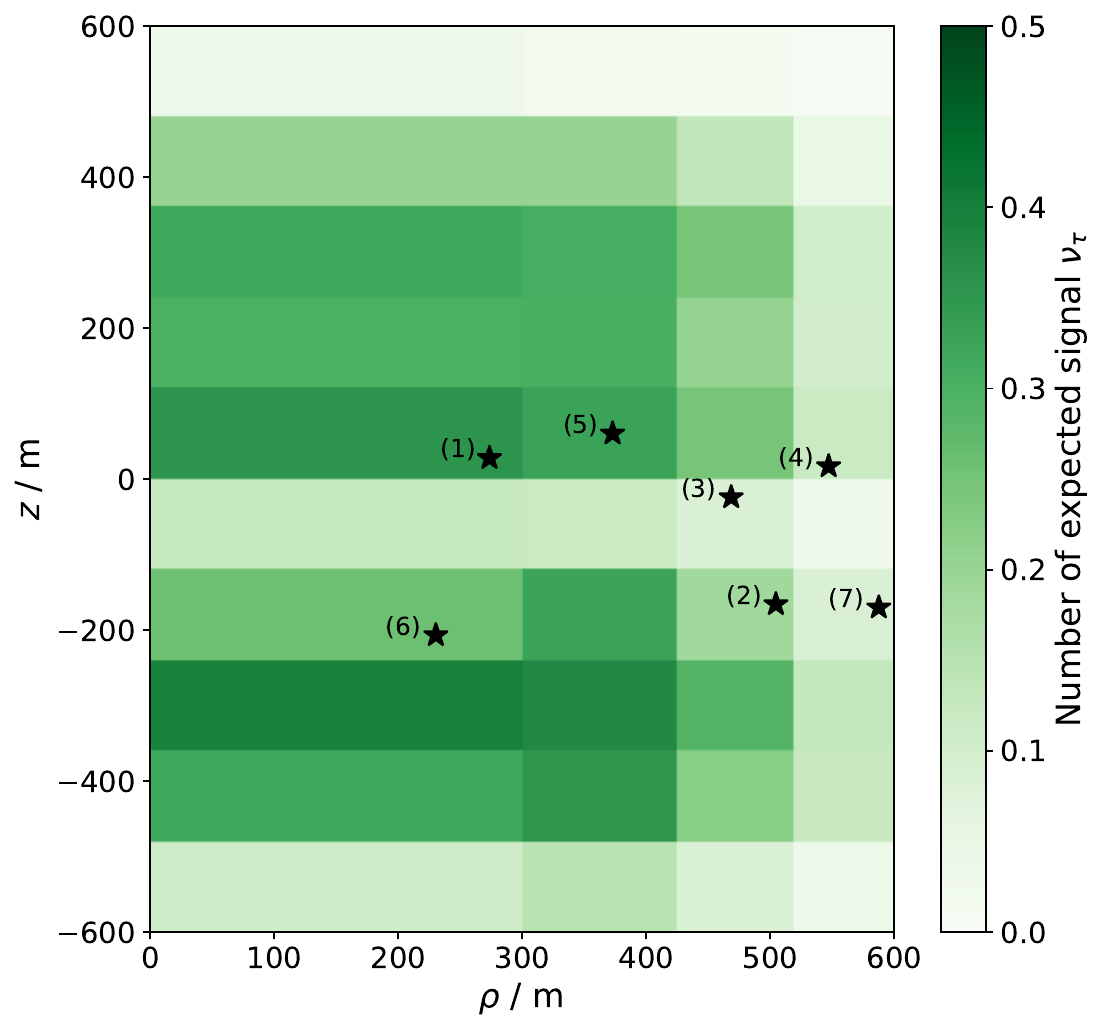}
    \caption{Side view of the reconstructed vertex positions~\protect\cite{IceCube:2021umt} of the seven $\nu_\tau^\mathrm{astro}$ candidate events, indicated by stars, as a function of vertical position $z$ and horizontal position $\rho \equiv \sqrt{x^2 + y^2}$. (Strings at the detector's edge have $\rho$ values from roughly $425-600~\mathrm{m}$.) The coordinates are measured with respect to the center of IceCube at a depth of 1950~m. The expected distribution of $\nu_{\tau}^\mathrm{astro}$ in 9.7 years for the astrophysical flux in Ref.~\protect\cite{IceCube:2015gsk} is overlaid.  (The numbers near each data point correspond to the events; see text. The horizontal bin widths are scaled with $\rho^2$.)}
    \label{fig:z_vs_rho}
\end{minipage}

\end{figure}

The seven $\nu_\tau^\mathrm{astro}$ candidate events were reconstructed \emph{a posteriori} using the algorithm described in Ref.~\cite{IceCube:2021umt}. We also applied the same reconstruction to the $\nu_{\tau,\,\mathrm{CC}}^\mathrm{astro}$ signal simulation.  Figure~\ref{fig:HullDistances} shows a top view of the reconstructed vertices for the seven events and Fig.~\ref{fig:z_vs_rho} shows a side view of the events, superimposed on the expected distribution from simulated signal. 
The numbers 1--7 next to each data point correspond to the images in Fig.~\ref{fig:SevenEventDisplays} (moving left to right and top to bottom).

From the depth distribution, it becomes apparent that the events seem to cluster close to a prominent, 100~m thick, dust layer centered around $z = -80$~m~\cite{IceCube:2013llx,Chirkin:2013lpu}. This region has high optical absorption, reducing detectable light. Photon scattering and absorption through all ice layers and their subsequent effects on the detection efficiency were included in our simulation. A Poisson goodness-of-fit test on the entire 2d histogram in Fig.~\ref{fig:z_vs_rho} gives a moderate $p$-value of 0.38, based on a suite of pseudo-trials. On the other hand, the projected $z$-distribution indicates a clustering that is inconsistent with expectation at the $3\sigma$ level, according to a Kuiper test~\cite{Kuiper}. 

To investigate further, we explored the possibility that the discrepancy might be due to mismodeling of the dust layer or due to atmospheric muons from prompt decays of charm or unflavored vector mesons. 

\subsection{Impact of the Dust Layer}

At the energies to which this analysis is sensitive, the selected events have $\tau$ leptons that travel 10--20~m.  At roughly 100~m thickness, the dust layer is too thick for it to enable background events to mimic signal events.  At our energies, the effect of the dust layer would be to obscure the light from one or both of the $\nu_\tau$ cascades, making the event look instead like a single cascade, or simply too dim, respectively.

Nevertheless, we performed several tests to verify that the CNNs were not unduly sensitive to light signals in the dust layer.  For candidate events near the dust layer, shifting the waveform times for all DOMs in the dust layer by $\pm 300$~ns, or even removing those DOMs entirely from the event, did not change the CNN response.  In another test, the times of individual pixels in the candidate events were shifted, and migration out of the signal region occurred only for shifts exceeding about 100~ns, well in excess of uncertainties expected from mismodeling of ice or DOM properties. (Each pixel holds the PMT charge in a 3.3~ns bin in the DOM's waveform.  The timing uncertainty due to ice properties is estimated to be about 20~ns for distances of about 100~m.) We also altered simulated background events near the dust layer and found that the CNN scores were similarly robust (see next section for more details). 

\subsection{Impact of Prompt Cosmic-Ray Muons}

The flux of muons from prompt decays of heavy mesons differs from the conventional flux as a function of both energy and arrival direction. To check whether the unanticipated $z$-distribution might be due to the (un-simulated) prompt muon component of cosmic-ray showers, a dedicated simulation was performed sampling muons from an $E^{-2}$ power-law spectrum and using a parameterized prompt model based on DPMJet-2.55~\cite{DPMJET255}. 
The targeted simulation was performed at depth, efficiently sampling muons from the parameterized model.  The simulated muon flux was then subjected to an analysis where we loosened the cut on the CNN $C_3$ score, designed to distinguish $\nu_\tau^\mathrm{astro}$ signal from muon tracks produced by $\nu_\mu$ interactions and downward-going muons, from 0.85 (0.95 with charge asymmetry requirements for outer strings) to 0.75. If the candidate data events were contaminated by muons, loosening $C_3$ as described would result in more simulated muons entering the signal region.  In this context, we performed a comparison of the spatial distribution across the detector of the generated muon events and that expected for data.
Since all of the simulated events that survive this CNN selection reconstruct as not fully contained, either near the outer edge or top of the detector, for this comparison we excluded the three candidate $\nu_\tau^\mathrm{astro}$ events that were well-contained within the fiducial region of IceCube. (We note that one of the remaining four candidate $\nu_\tau^\mathrm{astro}$ events in this subsample reconstructs as having passed through hundreds of meters of active detector volume above its interaction vertex, and the absence of appreciable light in this region further supports its neutrino hypothesis, but in what follows we do not incorporate this information.)

\begin{figure}[b]
    \centering
    \includegraphics[width=0.5\textwidth]{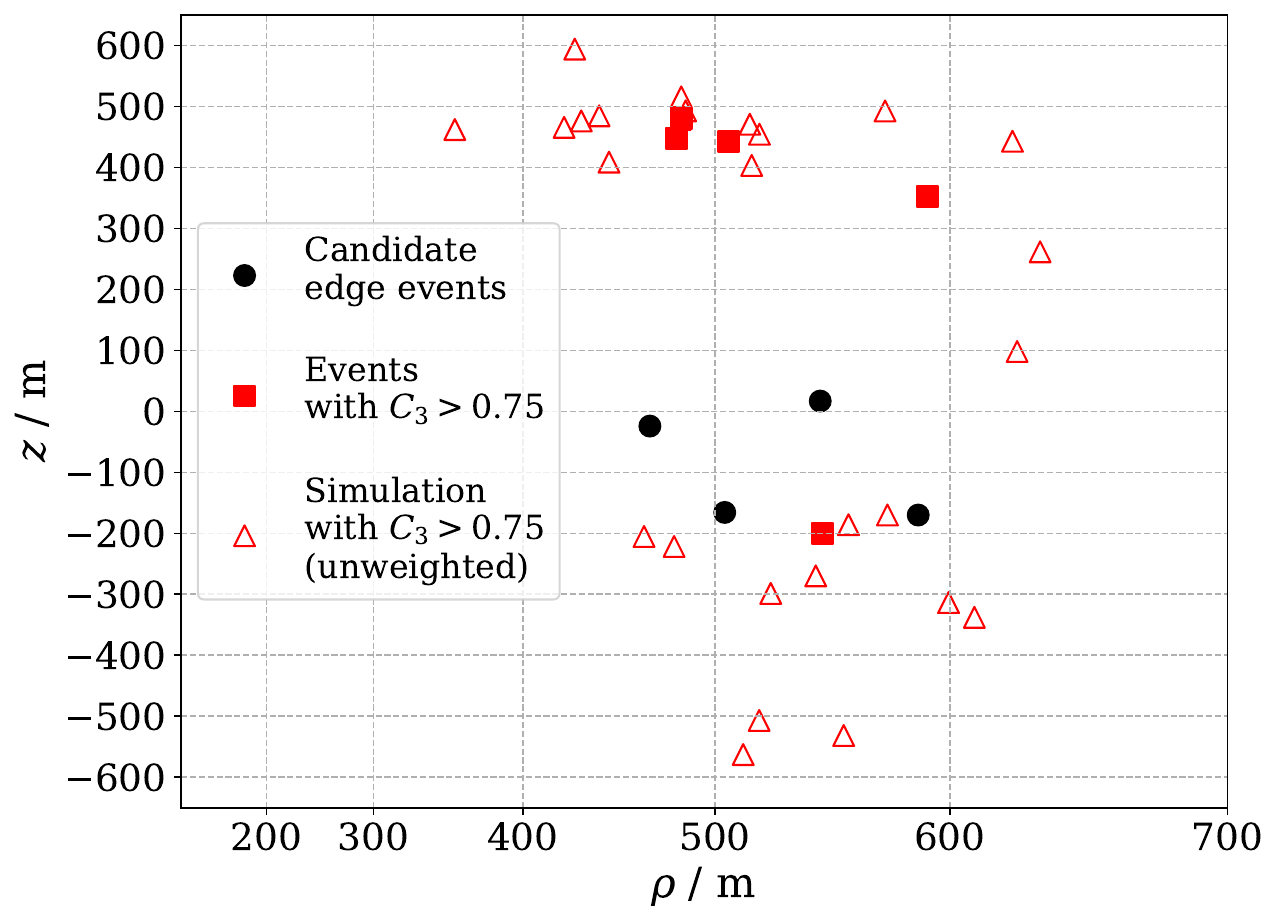}
    \caption{Distribution of $z$ vs. $\rho$ for
    candidate $\nu_\tau^\mathrm{astro}$ events classified as edge events (solid black circles), other data events satisfying a looser $C_3$ criterion (solid red squares; see text for details), and unweighted simulated downward-going muons also satisfying a looser $C_3$ (open red triangles).  All events were reconstructed using the algorithm described in Ref.~\protect\cite{IceCube:2021umt}.  (IceCube's use of event weighting in simulation is described in Ref.~\protect\cite{IceCube:2020tcq}. The horizontal axis is scaled with $\rho^2$.)}
    \label{fig:z_vs_rho_loosened_C3}
\end{figure}

Figure~\ref{fig:z_vs_rho_loosened_C3} shows a scatter plot of $z$ vs. $\rho$ $\equiv\sqrt{x^2 + y^2}$ (as measured from the center of IceCube at a depth of 1950~m) for 
data and unweighted simulated events classified as not fully contained. It is apparent that the additional data events allowed in by loosening the $C_3$ score mainly appear at the top of the detector. Comparing these data to the simulation of cosmic-ray muons, a Kolmogorov-Smirnov consistency test~\cite{KS} in $z$ gives a $p$-value of 0.1 for the hypothesis that all of the data events are muons.  Moreover, reverting to the original definition of the $C_3$ selection criterion leaves only eight unweighted simulated cosmic-ray muons, along with the four candidate edge events. Performing again a Kolmogorov-Smirnov consistency test in $z$ gives a $p$-value of 0.004, indicating that it is unlikely that all of the four candidate $\nu_\tau^\mathrm{astro}$ edge events are muons. 

Muons from cosmic-ray air showers enter primarily at the top of the detector. Muon rates were estimated from measurements of the cosmic-ray muon flux by IceCube and other instruments~\cite{IceCube:2015gsk,IceCube:2015wro,LVD:2019zlh,Bogdanov:2009ny}. These measurements are compatible with muons solely from $\pi^\pm/K^\pm$ decays, although upper bounds on the prompt contribution exist. We therefore did not include muons from charm decays in the background simulations used here. For the sake of completeness, however, we calculated the effect of arbitrarily increasing the atmospheric muon background by an order of magnitude with respect to the originally expected value of $0.005\pm 0.004$ (post-unblinding, using the sampling flux from Ref.~\cite{Gaisser:2011klf}). We find that the significance remains above $5\sigma$, even with such an increase. 

\subsection{Impact of Relaxing CNN scores $C_1$ and $C_2$}

\begin{figure}[h]
   \centering
   \includegraphics[width=0.5\columnwidth]{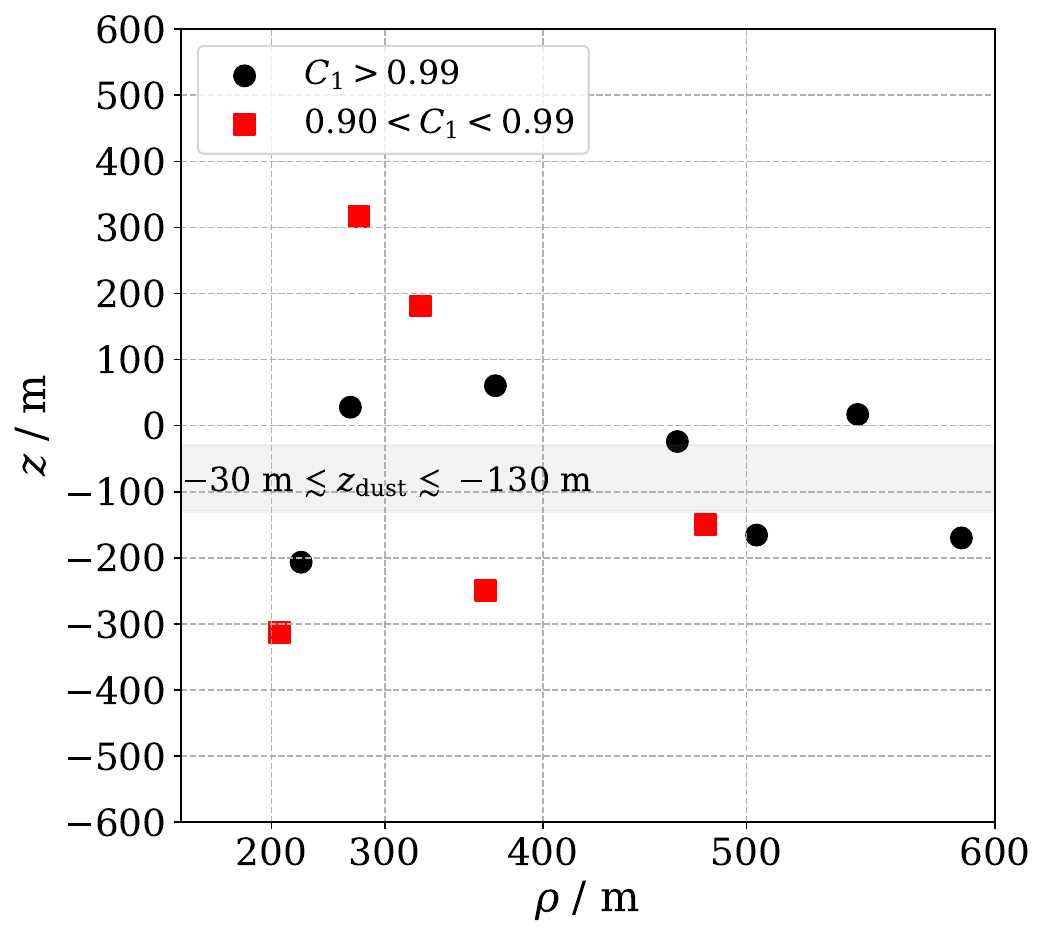}
   \caption{Distribution of event $z$ vs. $\rho \equiv \sqrt{x^2 + y^2})$, as measured from the center of IceCube at a depth of 1950~m, using the reconstruction in Ref.~\protect\cite{IceCube:2021umt} to estimate the event vertex position.  As discussed in the text, the original seven $\nu_\tau^\mathrm{astro}$ candidate events (black circles) appear to cluster near the prominent dust layer in the ice, shown as a horizontal gray band.  However, the five additional events (red squares) broaden the spatial distribution. All events were reconstructed using the algorithm described in Ref.~\protect\cite{IceCube:2021umt}. (The horizontal axis is scaled with $\rho^2$.)}
   \label{fig:rho_z_data_OldNew}
\end{figure}

Finally, we investigated the effects of less strict requirements on CNN scores $C_2$ and $C_1$. 
Loosening just the $C_2$ requirement only lets in events near the top of the detector, and therefore cannot provide an explanation for the observed $z$-clustering.
Figure~\ref{fig:rho_z_data_OldNew} shows a scatter plot of $z$ vs. $\rho \equiv\sqrt{x^2 + y^2}$, demonstrating the effect of loosening just the $C_1$ score criterion to $C_1 > 0.90$ to determine if a larger population of events is concentrated near the dust layer, the top of the detector, or at the detector perimeter, as would be expected for an enhanced background.  The additional events (red squares) instead broaden the spatial distribution, consistent with a statistical explanation for the originally observed clustering. From simulations, we expected that loosening $C_1$ would yield 9.4 signal and 2.9 background events, for a total of 12.3 events (assuming the IceCube GlobalFit~\cite{IceCube:2015gsk} flux), consistent with the 12 events observed. The additional five events have an average ``tauness'' $\langle P_\tau \rangle = 0.49$. The 12 events also exclude the null hypothesis at approximately $5\sigma$.

\section{CNN Robustness}

\subsection{Pre-Unblinding Data vs. Simulation Agreement}

\begin{figure}[h]
\centering
\includegraphics[width=0.9\columnwidth]{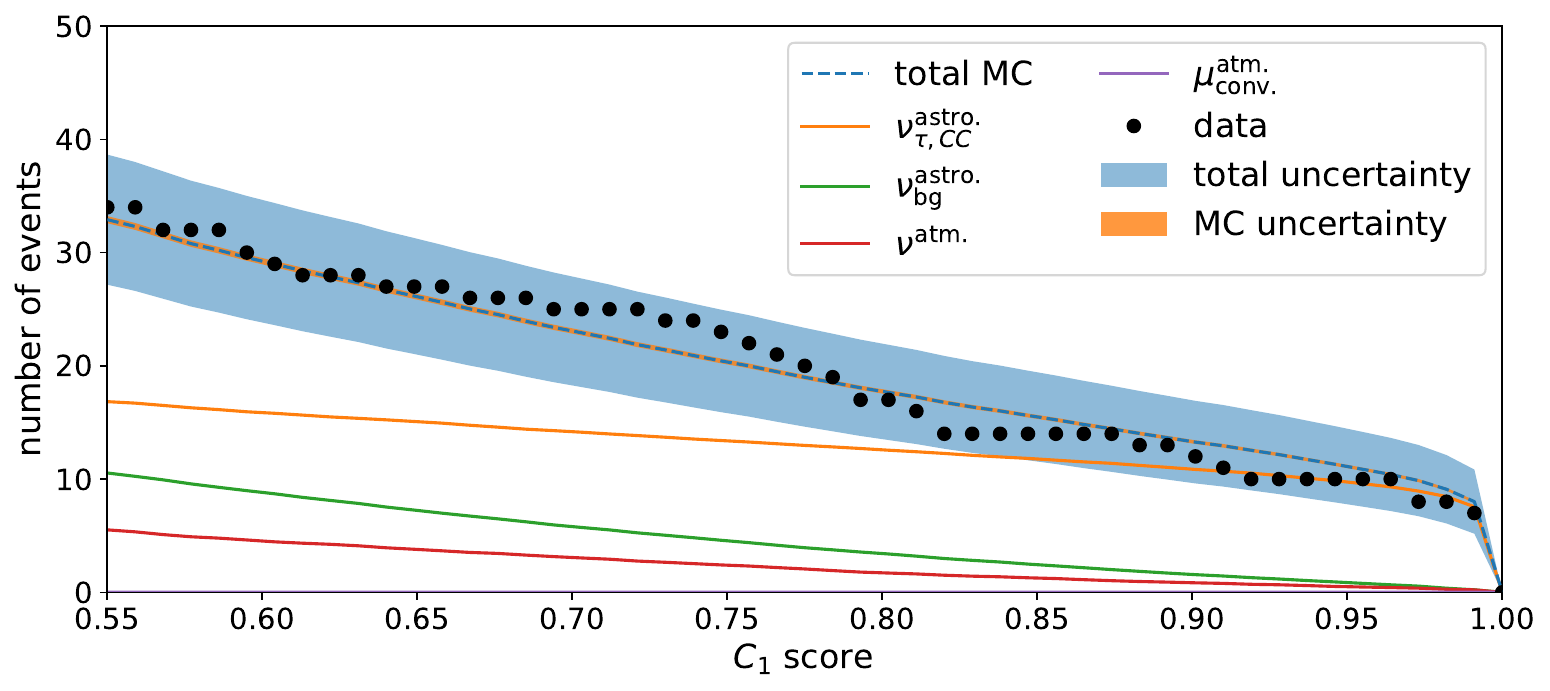}
\captionsetup{width=0.9\columnwidth}
\caption{Cumulative plot showing expected signal ($\nu_{\tau,CC}^\mathrm{astro.})$; expected backgrounds from other astrophysical neutrinos ($\nu_\mathrm{bg}^\mathrm{astro.}$), atmospheric neutrinos ($\nu^\mathrm{atm.}$), and conventional atmospheric muons ($\mu^\mathrm{atm.}_\mathrm{conv.}$); and observed data as a function of CNN score $C_1$.  The CNN scores $C_{2,3}$ were set to their final values.  Similar plots for $C_2$ and $C_3$ also show comparable agreement between data and simulation.
The IceCube GlobalFit $\nu^\mathrm{astro}$ flux~\protect\cite{IceCube:2015gsk} is assumed.}
\label{fig:C1_data_MC}
\end{figure}

Prior to unblinding we investigated the agreement in the CNN scores between data and simulation in the regions populated by background events. Figure~\ref{fig:C1_data_MC} shows a cumulative plot of the number of data and expected signal and background events vs. the CNN score $C_1$; CNN scores $C_{2,3}$ show similar levels of agreement.

\subsection{Post-Unblinding Tests}

Here we describe in more detail the various data-driven and simulation-based tests of CNN robustness that we performed. For the first suite of tests, we define the background region as $C_{1,2,3} < (0.9,0.9,0.75)$, comprising 8,175 of the original 8,188 events passing the preliminary selection criteria. We applied randomized scale factors to DOM waveforms that artificially increased or decreased the magnitude of the detected light level within expected systematic uncertainties, in five distinct patterns:
\begin{itemize}
    \item each of 180 DOMs were randomly scaled independent of one another,
    \item dividing the detector into regions in depth, the group of DOMs in each region was randomly scaled by the same factor, as follows:
    \begin{itemize}
        \item 20 groups of 9 DOMs each in regions in depth of about 50~m,
        \item 15 groups of 12 DOMs each in regions in depth of about 68~m,
        \item 12 groups of 15 DOMs each in regions in depth of about 85~m, and
    \end{itemize}
    \item on each of the two less-illuminated strings, all 60 DOMs were randomly scaled by the same factor (a total of two distinct factors were used).
\end{itemize}
The first pattern addresses relative DOM detection efficiencies~\cite{IceCube:2016zyt}, the next three address ice optical properties as a function of depth, and the fifth addresses ice birefringence~\cite{chirkin2019light} as a function of azimuthal angle.

For each pattern, we performed 750 trials per event, for a total of $(5 \times 8175 \times 750)$ or about $3\cdot 10^7$ trials.  We found that the probability of background-to-signal migration did not exceed $(2 \pm 0.2)\cdot 10^{-5}$ in any of the tests, corresponding to $0.16 \pm 0.02$ events, and that migration occurred only when events were already close to the signal region.  The final significance remains above $5\sigma$ whether we use simulated data sets (described earlier) or this data-driven approach to handle these detector systematics.  The same tests performed on the seven signal events showed a signal-to-background migration probability of $(3 \pm 0.8) \cdot 10^{-3}$.

We also employed targeted tests to estimate the CNNs' robustness against less likely changes in the underlying raw data. In candidate events with prominent double pulse waveforms, we interpolated between the two peaks to merge together the first and second pulses, and found that this did not cause any candidate events to migrate out of the signal region.  As mentioned earlier, shifting pixel arrival times in individual DOM waveforms in candidate events by up to 100~ns did not appreciably change the CNN response.   

We applied adversarial attacks~\cite{moosavi2016deepfool} against the candidate events, using an optimization algorithm to find the pixel(s) whose physically reasonable alterations resulted in the largest changes to the CNN scores.  Just one of the seven candidate events could be forced to migrate, and only when the average change over all pixels was at least 2.5\%, a situation that is well outside our estimated uncertainties.  Similarly attacking simulated background events, we found that in no particular region of the detector did the CNNs exhibit heightened susceptibility, and generally the changes required to induce migration were much larger than allowed by our uncertainties.  We also attacked 634 simulated astrophysical $\nu_e$, allowing the individual pixel uncertainties to be as high as 10\%, finding in this harsh test that only one simulated $\nu_e$ was misclassified as a $\nu_\tau$.  Finally, we attacked the candidate events after randomly varying their pixel values with 10\% uncertainty.  Using $10^4$ trials per event, only one event was found to have a $(2.1 \pm 0.14)\%$ migration probability. These targeted tests, and other studies described earlier in this Supplemental Material, indicate that even under quite harsh conditions, the CNNs remained capable of rejecting the background events while retaining the candidate signal events.

\end{document}